\documentclass[journal]{IEEEtran}

\ifCLASSINFOpdf
  \usepackage[pdftex]{graphicx}
  \DeclareGraphicsExtensions{.pdf,.jpeg,.png}
\else
  \usepackage[dvips]{graphicx}
  \DeclareGraphicsExtensions{.eps}
\fi

\usepackage{amsmath,amsfonts,amssymb}
\usepackage{cite}
\usepackage{color}
\usepackage{pstricks,pst-plot}
\usepackage{pst-sigsys}
\usepackage{psfrag}
\usepackage{ifthen}
\usepackage{balance}
\usepackage{multirow}
\usepackage{pbox}
\usepackage[caption=false,font=footnotesize]{subfig}

\usepackage{mathrsfs}
\usepackage{amsbsy}

\newcommand{\vu}[2]{\mbox{$#1\,\text{#2}$}}

\definecolor{ForestGreen}{rgb}{0.2, 0.6, 0.1}
\definecolor{Grey}{rgb}{0.5, 0.5, 0.5}
\definecolor{Black}{rgb}{0, 0, 0}

\newcommand{\Erik}[1]{\textcolor{Black}{#1}}

\newcommand{\E}[2]{\mathbb{E}_{#1}\left\{ #2 \right\}}   
\newcommand{\tr}{\mathrm{tr}}                       
\newcommand{\conv}{\ast}                            

\newcommand{\anchoridxsym}{j}               
\newcommand{\anchoridx}{^{(\anchoridxsym)}} 
\newcommand{\anchoridxub}{J}                

\newcommand{\agentidxsym}{m}               
\newcommand{\agentidx}{^{(\agentidxsym)}}  
\newcommand{\agentidxub}{M}                

\newcommand{\anchoragentidx}{^{(\anchoridxsym,\agentidxsym)}}  
\newcommand{\agentagentidx}{^{(\agentidxsym,\agentidxsym)}}  	 

\newcommand{\mpcidxsym}{k}          
\newcommand{\mpcidx}{_{\mpcidxsym}} 
\newcommand{\mpcidxub}{K}           

\newcommand{\timestepsymub}{N}          

\newcommand{\vaconstridxsym}{q}         				
\newcommand{\vaconstridx}{_{\vaconstridxsym}}   

\newcommand{\jnum}[1]{^{(#1)}}	
\newcommand{\jmnum}[2]{^{(#1,#2)}}	
\newcommand{\knum}[1]{_{#1}}		

\newcommand{\q}{\vaconstridx}

\newcommand{\arbsymb}{\xi} 			
\newcommand{\arb}{\jnum{\arbsymb}}

\newcommand{\arbsymbtwo}{\eta} 			
\newcommand{\arbtwo}{\jnum{\arbsymbtwo}}

\newcommand{\f}[2]{{\frac{#1}{#2}}}				
\newcommand{\pa}{\partial}								
\newcommand{\fp}[2]{{\f{\pa #1}{\pa #2}}}	

\newcommand{\eff}{\bar{\zeta}}						
\newcommand{\flipmtx}{\left[\begin{array}{rr}1&0\\0&-1\end{array} \right]}

\newcommand{\GT}{{\bf H}_\text{Ag}} 
\newcommand{\GR}{{\bf H}_\text{An}} 
\newcommand{\GM}{{\bf H}_\text{Mo}} 

\newcommand{\clockoff}{\epsilon}
\newcommand{\clockoffvec}{\boldsymbol{\epsilon}}
\newcommand{\clockoffinflmat}{\pmb{\mathcal{C}}}
\newcommand{\clockoffinfleigvec}{{\bf c}}
\newcommand{\IDMat}{{\bf I}}
\newcommand{\JacMat}{{\bf J}}
\newcommand{\FIM}{\pmb{\mathcal{I}}}
\newcommand{\FIMAg}{\pmb{\mathcal{I}}_\mathrm{Ag}}
\newcommand{\FIMAn}{\pmb{\mathcal{I}}_\mathrm{An}}
\newcommand{\FIMMo}{\pmb{\mathcal{I}}_\mathrm{Mo}}
\newcommand{\FIMC}{\pmb{\mathcal{I}}_\mathrm{C}}

\newcommand{\bd}{{\bf d}}
\newcommand{\bp}{{\bf p}}

\newcommand{\br}{{\bf r}}

\newcommand{\be}{{\bf e}}
\newcommand{\bh}{{\bf h}}

\newcommand{\bs}{{\bf s}}
\newcommand{\bv}{{\bf v}}

\newcommand{\bn}{{\bf n}}

\newcommand{\bb}{{\bf b}}
\newcommand{\bl}{{\bf l}}

\newcommand{\bH}{{\bf H}}
\newcommand{\bR}{{\bf R}}
\newcommand{\bI}{{\bf I}}

\newcommand{\bT}{{\bf T}}
\newcommand{\bO}{{\bf 0}}

\newcommand{\bP}{{\bf P}}

\newcommand{\bA}{{\bf A}}
\newcommand{\bB}{{\bf B}}
\newcommand{\bS}{{\bf S}}
\newcommand{\bC}{{\bf C}}
\newcommand{\bD}{{\bf D}}

\newcommand{\bM}{{\bf M}}
\newcommand{\bF}{{\bf F}}
\newcommand{\bW}{{\bf W}}
\newcommand{\bZero}{{\bf 0}}
\newcommand{\bL}{{\bf L}}
\newcommand{\bG}{{\bf G}}

\newcommand{\balpha}{\boldsymbol{\alpha}}
\newcommand{\btheta}{{\boldsymbol{\theta}}}
\newcommand{\bTheta}{{\boldsymbol{\Theta}}}
\newcommand{\bpsi}{{\boldsymbol{\psi}}}
\newcommand{\bPsi}{{\boldsymbol{\Psi}}}
\newcommand{\btau}{\boldsymbol{\tau}}

\newcommand{\bLambda}{\boldsymbol{\Lambda}}

\newcommand{\bLA}{\boldsymbol{\Lambda}_\mathrm{A}}
\newcommand{\bLB}{\boldsymbol{\Lambda}_\mathrm{B}}
\newcommand{\bLC}{\boldsymbol{\Lambda}_\mathrm{C}}

\newcommand{\Ts}{{T_\mathrm{s}}}

\newcommand{\Tp}{{T_\mathrm{p}}}
\newcommand{\fc}{{f_\mathrm{c}}}

\newcommand{\bCn}{{\bf C}_\mathrm{n}}

\newcommand{\bCc}{{\bf C}_\mathrm{c}}

\newcommand{\sigman}{\sigma_\mathrm{n}}

\newcommand{\diff}{\mathrm{diff}}
\newcommand{\rolloff}{R}

\newcommand{\SINR}{\mathrm{SINR}}

\newcommand{\T}{^\mathrm{T}}
\newcommand{\R}{^\mathrm{R}}
\newcommand{\I}{^\mathrm{I}}
\newcommand{\Herm}{^\mathrm{H}}
\newcommand{\diag}{\mathrm{diag}}

\renewcommand{\mod}[1]{_{\mathrm{mod\,}#1}}

\newcommand{\pind}[1]{\mathbf{p}_{\mathrm{#1}}}   

\definecolor{Pblue}{rgb}{0, 0.5, 1}
\definecolor{Fgreen}{rgb}{0.2, 0.6, 0.1}
\definecolor{darkgrey}{rgb}{0.3, 0.3, 0.3}

\begin{document}
\title{Evaluation of Position-related Information in Multipath Components for Indoor Positioning}
\author{Erik Leitinger,~\IEEEmembership{Student Member,~IEEE}, Paul Meissner,~\IEEEmembership{Member,~IEEE}, Christoph R\"udisser, \\Gregor Dumphart, and Klaus Witrisal,~\IEEEmembership{Member,~IEEE}
\thanks{E. Leitinger, P. Meissner, C. R\"udisser, and K. Witrisal are with Graz University of Technology, Graz, Austria, email: \{erik.leitinger, paul.meissner, witrisal\}@tugraz.at}
\thanks{G. Dumphart is with ETH Zurich, Zurich, Switzerland, email: dumphart@nari.ee.ethz.ch}
\thanks{Manuscript received Aug. 14, 2014; revised Dec. 6, 2014 and Jan. 23, 2015, accepted Feb. 16, 2015.}
\thanks{This work was supported by the Austrian Science Fund (FWF), National Research Network SISE, Project S10610 and by the Austrian Research Promotion Agency (FFG), 
KIRAS PL3, grant no. 832335 ``LOBSTER''.}}

\maketitle

\begin{abstract}
  Location awareness is a key factor for a wealth of wireless indoor applications. Its provision requires the careful
fusion of diverse information sources. For agents that use radio signals for localization, this information may either
come from signal transmissions with respect to fixed anchors, from cooperative transmissions inbetween agents, or from
radar-like monostatic transmissions. Using a-priori knowledge of a floor plan of the environment, specular multipath
components can be exploited, based on a geometric-stochastic channel model. In this paper, a unified framework is
presented for the quantification of this type of position-related information, using the concept of equivalent Fisher
information. We derive analytical results for the Cram\'er-Rao lower bound of multipath-assisted positioning,
considering bistatic transmissions between agents and fixed anchors, monostatic transmissions from agents,  cooperative
measurements inbetween agents, and combinations thereof, including the effect of clock offsets. Awareness of this
information enables highly accurate and robust indoor positioning. Computational results show the applicability of the framework for the characterization of the localization capabilities of a given environment, quantifying the influence of different system setups, signal parameters,  and the impact of path overlap.

\end{abstract}


\section{Introduction}
\label{sec:Introduction}
\graphicspath{{./figures/final_fig/}}

Location awareness is a key component of many future wireless applications. Achieving the needed level of 
accuracy \emph{robustly}\footnote{We define robustness as the percentage of cases in which a system can achieve
its given potential accuracy.} is still elusive, especially in indoor environments which are
characterized by harsh multipath conditions. Promising candidate systems thus either use sensing technologies that
provide remedies against multipath or they fuse information from multiple information sources
\cite{ShenJSAC2012,Conti2014}. WLAN-based systems make use of existing infrastructure and exploit the position
dependence of the received signal strength \cite{MazuelasSTSP2009}. However, the latter shows a relatively large
variance w.r.t. the position-related parameters such as the distance, even with an optimized deployment
\cite{FiccoTMC2014}.

In \emph{Multipath-assisted indoor positioning,} multipath components (MPCs) can be associated to the local
geometry using a known floor plan. In this way, MPCs can be seen as signals from additional (virtual) anchors (VAs).
Ultra-wideband (UWB) signals are used because of their superior time resolution and to facilitate the separation of
MPCs. Hence, additional position-related information is exploited that is contained in the radio signals.

This is in contrast to competing approaches, which either detect and avoid non-line-of-sight (NLOS) measurements
\cite{MaranoJSAC2010}, mitigate errors induced by strong multipath conditions \cite{WymeerschTC2012}, or employ more
realistic statistical models for the distribution of the range estimates \cite{LuICC2013}. Cooperation between agents is
another method to increase the amount of available information \cite{WymeerschProc2009} and thus to reduce the
localization outage. Actual exploitation of multipath propagation requires prior knowledge \cite{ShenGlobecomm2009}.
This can be the floor plan, like in this work and related approaches \cite{ParhizkarICASSP2014}, or a set of known
antenna locations to enable beamforming (e.g. in imaging \cite{LeigsneringSPM2014}). In an inverse problem, the room
geometry can be inferred from the multipath and known measurement locations \cite{DokmanicPNAS2013}.

Insight on the position-related information that is conveyed in the signals \cite{MeissnerEUCAP2012} can be gained by an
analysis of performance bounds, such as the Cram\'er-Rao lower bound (CRLB), which is the lower bound of the
covariance matrix of an unbiased estimator for a vector parameter. Using the concept of equivalent Fisher information
matrices (EFIMs) \cite{ShenTIT2010,ShenTIT2010A}, allows for analytic evaluation of the CRLB by blockwise inversion of
the Fisher information matrix (FIM) \cite{Qi2006,GodrichTIT2010}. 

A proper channel model is paramount to capture the information contained in MPCs. 
%
\begin{figure}[t]
  \centering
  \includegraphics[width=1\columnwidth]{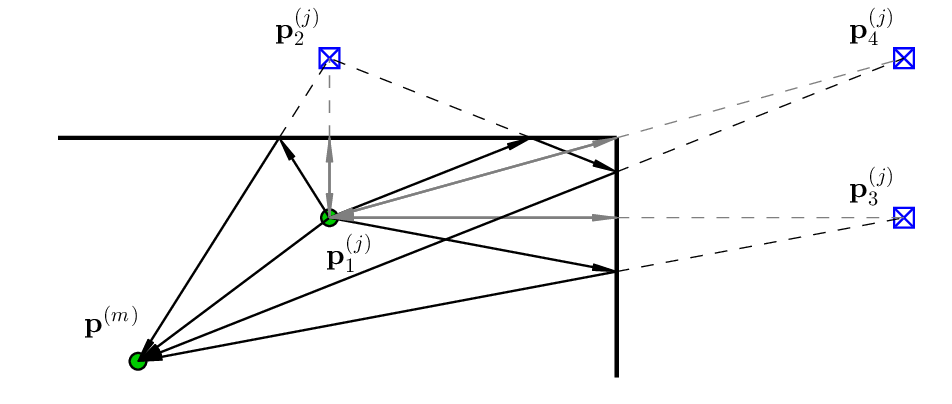}
  \caption{Illustration of multipath geometry using VAs for (i) bistatic transmissions (black) between an anchor at
  $\pind{1}\anchoridx$ and an agent at $\pind{}\agentidx$ and for (ii) a monostatic  measurement (gray) by an
  agent at $\pind{1}\anchoridx$. } 
  \label{fig:VAgeometry}
\end{figure}
It is common \cite{RichterVTC2005, MichelusiTSP2012, DecarliJSTSP2014,
SantosTWC2010a, KaredalTWC2007} to differentiate between resolvable MPCs which origin from specular
reflections or scatterers and so-called dense or diffuse multipath (DM), which comprises all other ``energy
producing'' components that can not be resolved by the measurement aperture. This part of the channel is often modeled
statistically since many unresolvable components add up in one delay bin of the channel impulse response. An established
approach to describe these statistics is to use parametric models for the power delay profile (PDP)
\cite{RichterVTC2005,MichelusiTSP2012}. The overall models are often referred to as hybrid geometric-stochastic channel
models (GSCMs). For the analysis presented in this paper, propagation effects other than the geometrically modeled MPCs
constitute interference to useful position-related information. This interference is also called diffuse multipath (DM)
\cite{WitrisalICC2012} and modeled as a colored noise process with non-stationary statistic.

Fig.~\ref{fig:VAgeometry} illustrates the geometric model for multipath-assisted positioning. A signal exchanged
between an anchor at position $\pind{1}\anchoridx$ and an agent at $\pind{}\agentidx$ contains specular reflections at
the room walls, indicated by the black lines.\footnote{Since the radio channel is reciprocal, the assignment of
transmitter and receiver roles to anchors and agents is arbitrary and this choice can be
made according to higher-level considerations.} These reflections can be modeled geometrically using VAs
$\bp\mpcidx\anchoridx$, mirror images of the anchor w.r.t. walls that can be computed from the floor
plan \cite{Meissner2014a,Borish1984,Kunisch2003}. We call this the \emph{bistatic} setup, where the fixed anchors and
the floor plan constitute the available infrastructure. In a \emph{cooperative} setup, agents localize themselves using
bistatic measurements inbetween them. Here, the node at $\pind{1}\anchoridx$ is an 
agent that plays the role of an anchor (and thus provides a set of VAs) for the agent at $\pind{}\agentidx$. 
If the agents are equipped accordingly, they can use \emph{monostatic} measurements, indicated by the gray lines. Here,
the node at $\pind{1}\anchoridx$ acts as anchor for itself with its own set of VAs.

For these measurement setups, we analyze the following scenarios isolated to get insights on different
effects of interest: (i) Multipath-Sync with known clock-offset between anchors and agents, (ii)
Multipath-NSync with unknown clock-offset between anchors and agents and optionally also between the individual
anchors, and (iii) Multipath-Coop with cooperation between the agents, monostatic measurements, and possibly additional
fixed anchors. Clock-synchronization for impulse radio UWB has shown to achieve a synchronization accuracy in order of
\vu{1}{ns}, which results still in large localization errors \cite{CarboneJIM2013}. As a consequence, we estimate the
clock-offset jointly, solely based on the received signal and the a-priori known floor plan. Only the differences
between the arrival times of MPCs carry position-related information in this case, not the time of arrival as in the
synchronized one.

For a tracking application, we have coined the terms \emph{multipath-assisted indoor navigation and tracking (MINT)} for
the bistatic setup \cite{WitrisalICC2012}, and Co-MINT \cite{FroehleICC2013} for the cooperative setup. The robustness
and accuracy of MINT have been reported in \cite{Meissner2014a,MeissnerICC2014,LeitingerICC2014} and references therein.
Also, a real-time demonstration system has been realized \cite{MeissnerICC2014}.

The key contributions of this paper are:
\begin{itemize}
  \item We present a mathematical framework for the quantification of position-related information contained in geometrically modeled specular reflections in (ultra) wideband wireless signals under DM.
  \item This information is quantified for conventional bistatic, monostatic, and cooperative measurement scenarios, optionally including unknown clock offsets, allowing for important insights that can be used in the design of a localization system.
  \item The results show the relevance of a site-specific, position-related channel model for indoor positioning and
  the components it comprises of. This position-related FIM is a measure for \emph{accuracy} and as a
  further consequence, it can also be seen as indicator for \emph{robustness}, since it increases with the number
  of useful MPCs, which also makes algorithms based on multipath-assisted approach more \emph{robust}.
  \item We validate, using real measurements, the usefulness of the derived bounds and of the introduced
  signal-to-interference-plus-noise-ratio (SINR) as a measure for position-related information.
\end{itemize}


The paper is organized as follows: Section \ref{sec:model} introduces the geometric-stochastic signal
model that is used in Section \ref{sec:CRLB_channel} to derive the CRLB on the position estimation error. Section
\ref{sec:geometry} describes the relationship between signal parameters and node positions in a generic form. These
results are used in Section \ref{sec:CRLB_PE} to derive the CRLB for the different scenarios. Finally, Sections
\ref{sec:results} and \ref{sec:concl} wrap up the paper with results, discussions, and conclusions.

\textit{Mathematical notations}: 
$\E{z}{\cdot}$ represents the expectation operator with respect to the random variable $z$. $[\bA]_{n,m}$ is the
$(n,m)$-th element of matrix $\bA$; $\bA_{N \times M}$ indicates the size of a matrix. $\|\cdot\|$ is the Euclidean
norm, $|\cdot|$ is the absolute value, and $(\ast)$ denotes convolution. $\bA \succeq \bB$ means that $\bA - \bB$ is
positive semidefinite.  $\IDMat_\timestepsymub$ is the identity matrix of size $\timestepsymub$. $(\cdot{•})\Herm$
is the Hermitian conjugate. $\tr\{ \cdot \}$ and $\mathrm{diag}\{ \cdot \}$ are the trace and the diagonal of a square
matrix, respectively.

\section{Signal Model}
\label{sec:model}
In Sections~\ref{sec:model} and \ref{sec:CRLB_channel}, we simplify the setup---for the ease of readability---to a
single (fixed) anchor located at position $\bp_1\in \mathbb{R}^2$ and one agent at position $\bp \in \mathbb{R}^2$. Note
that two-dimensional position coordinates are used throughout the paper, for the sake of
simplicity\footnote{The extension to three dimensional coordinates is straightforward.}. A \emph{baseband}
UWB signal $s(t) $ is exchanged between the anchor and the agent. The corresponding received signal is modeled as
\cite{WitrisalICC2012}
\begin{align}\label{eq:rx_signal}
  r(t) &= r_{\det}(t) + r_{\diff}(t) + w(t) \nonumber \\
  &= \sum_{\mpcidxsym=1}^{\mpcidxub}\alpha\mpcidx s(t-\tau\mpcidx) + (s\conv\nu)(t - \clockoff) + w(t).
\end{align}
The first term $r_{\det}(t)$ describes a sum of $K$ deterministic MPCs with complex amplitudes $\{\alpha_{k}\}$ and
delays $\{\tau_{k}\}$. We model these delays by VAs at positions $\bp_k \in \mathbb{R}^2$, 
yielding $\tau_{k} = \frac{1}{c}\|\bp - \bp_{k}\| + \clockoff$, with $k = 1 \ldots K$, where $c$ is the speed of light
and $\clockoff$ represents the clock-offset due to clock asynchronism. $K$ is equivalent to the number of visible VAs at
the agent position $\bp$ \cite{Meissner2014a}. We assume the energy of $s(t)$ is normalized to one. 

The second term $r_{\diff}(t)$ denotes the convolution of the transmitted signal $s(t)$ with the DM $\nu(t)$, which is
modeled as a zero-mean Gaussian random process. Note that the statistic of $r_{\diff}(t)$ is non-stationary in the
delay domain and it is colored due to the spectrum of $s(t)$. For DM we assume uncorrelated scattering along the delay
axis $\tau$, hence the auto-correlation function (ACF) of $ \nu(t)$ is given by
\begin{equation}
 K_{\nu}(\tau,u)= \E{\nu}{\nu(\tau)[\nu(u)]^*} = S_{\nu}(\tau)\delta(\tau-u),
\end{equation}
where $S_{\nu}(\tau)$ is the PDP of DM at the agent position $\bp$. The DM process is assumed to
be quasi-stationary in the  spatial domain, which means that $S_{\nu}(\tau)$ does not change in the vicinity of position
$\bp$ \cite{MolischTPROC2009}. The PDP $S_{\nu}(\tau)$ is crucial to represent the power ratio between useful
deterministic MPCs and DM (along the delay axis $\tau$) and it is represented by an arbitrary function which can be
estimated from an ensemble of measurements \cite{MichelusiTSP2012} \footnote{The PDP for instance can be estimated
globally for an anchor placed in a room from sets of measurements distributed over the according floor plan and then it
can be updated during tracking of an agent \cite{Meissner2014a}.}, rather than a parametric PDP
\cite{RichterVTC2005}. We will assume that the DM statistic is known a-priori to be able to analyze the influence of DM
on the CRLB in closed form, with no parametric restriction on the DM PDP  $S_{\nu}(\tau)$. With this, our results will
show that information coming from MPCs is quantified by a signal-to-interference-plus-noise-ratio (SINR) for these MPCs,
which represents the power ratio between useful deterministic MPC and impairing DM plus noise. Finally, the last term
$w(t)$ denotes an additive white Gaussian noise (AWGN) process with double-sided power spectral density (PSD) of
$N_0/2$. 

In the following, we will drop the clock-offset $\epsilon$. We will re-introduce it in Section \ref{sec:CRLB_TDOA} where
the Multipath-NSync setup is studied.

\section{Cram{\'e}r-Rao Lower Bound}
\label{sec:CRLB_channel}
The goal of multipath-assisted indoor positioning is to estimate the agent's position $\bp$ from the signal
waveform (\ref{eq:rx_signal}), exploiting the knowledge of the VA positions $\{\bp_k\}$, in presence of diffuse
multipath and AWGN with known statistics. Let $\hat{\btheta}$ denote the estimate of the position-related parameter
vector $\btheta  = [\bp\T \ \Re\balpha\T \ \Im\balpha\T]\T \in \mathbb{R}^{D_\btheta}$, where $\Re\balpha =
[\Re\alpha_{1},\dots,\Re\alpha_{K}]\T$ and $\Im\balpha = [\Im\alpha_{1},\dots,\Im\alpha_{K}]\T$ are the real and
imaginary parts of the complex amplitudes $\balpha$, respectively, which are nuisance parameters. According to the
information inequality, the error covariance matrix of $\btheta$ is bounded by \cite{Kay1993}
\begin{equation}
  \E{\br|\btheta}{\big(\hat{\btheta}-\btheta \big)\big(\hat{\btheta}-\btheta \big)\Herm} \succeq \FIM_\btheta^{-1},
\end{equation}
where $\FIM_\btheta \in \mathbb{R}^{D_\btheta \times D_\btheta}$ is the Fisher information matrix
(FIM) and its inverse represents the CRLB of $\btheta$. We apply the chain rule to derive this CRLB 
(cf. \cite{ShenTIT2010, GodrichTIT2010}), i.e., the FIM $\FIM_\btheta$ is computed from the FIM of the signal 
parameter vector $\bpsi = \big[\boldsymbol{\tau}\T , \Re\balpha\T , \Im\balpha\T \big]\T \in \mathbb{R}^{D_\bpsi}$,
where $\btau = [\tau_{1},\dots,\tau_{K}]\T$ represents the vector of position-related delays. We get
\begin{equation}\label{eq:ParmTrans}
 \FIM_\btheta = \JacMat\T \FIM_\bpsi \JacMat
\end{equation}
with the Jacobian 
\begin{align}\label{eq:JacobianGenereal}
  \JacMat &= \frac{\partial\bpsi}{\partial\btheta} \in \mathbb{R}^{D_\bpsi \times D_\btheta}.
\end{align}
The FIM $\FIM_\bpsi \in \mathbb{R}^{D_\bpsi \times D_\bpsi }$ of the signal model parameters can be computed from the
likelihood function $f(\br|\bpsi)$ of the received signal $\br$ conditioned on parameter vector $\bpsi$,
\begin{align}\label{eq:FIM_def}
  \FIM_\bpsi &= \E{\br|\bpsi}
  {\left[\frac{\partial}{\partial\bpsi} \ln f(\br|\bpsi) \right]
  \left[\frac{\partial}{\partial\bpsi} \ln f(\br|\bpsi) \right]\T}. 
\end{align}

\subsection{Likelihood Function}
\label{sec:LHF}
The likelihood function $f(\br|\bpsi)$ is defined for the sampled received signal vector 
$\br= [r(0),r(\Ts),\dots,r((\timestepsymub-1) \Ts)]\T$ $\in \mathbb{C}^\timestepsymub $, containing $\timestepsymub$
samples at rate $1/\Ts$. Using the assumption that AWGN and DM are both Gaussian, it is given by
\begin{align}\label{eq:sampledLHF}
  f(\br|\bpsi) & \propto \exp \left\{-( \br - \bS\balpha)\Herm\bCn^{-1}( \br -
  \bS\balpha )\right\} \nonumber \\
  & \propto \exp \left\{2\Re \left\{ \br\Herm\bCn^{-1}\bS\balpha \right\}-
  \balpha\Herm\bS\Herm\bCn^{-1}\bS\balpha \right\}
\end{align}
where  $\bS= [\bs_{\tau_1}, \dots,\bs_{\tau_K}]\in\mathbb{R}^{\timestepsymub \times K}$ is the signal matrix 
%
%
containing delayed versions $\bs_{\tau\mpcidx}=[s(-\tau\mpcidx),s(\Ts-\tau\mpcidx),\dots,s((\timestepsymub-1)
\Ts-\tau\mpcidx)]\T$ of the sampled transmit pulse and $\bCn = \sigman^2\IDMat_{\timestepsymub} + \bCc
\in\mathbb{R}^{\timestepsymub \times \timestepsymub}$ denotes the co-variance matrix of the noise processes. The vector
of AWGN samples has variance $\sigman^2 = N_0/\Ts$; the elements of the DM co-variance matrix are given by
$[\bCc]_{n,m} = \Ts\sum_{i=0}^{\timestepsymub-1} S_\nu(i\Ts)s(n\Ts-i\Ts)s(m\Ts-i\Ts)$ (see Appendix~\ref{app:lhf}).

\subsection{FIM for the Signal Model Parameters}
\label{sec:CRLB_SigMod}

\subsubsection{General Case}
The FIM $\FIM_\bpsi$ is obtained from (\ref{eq:FIM_def}) with (\ref{eq:sampledLHF}). Following the notation of
\cite{ShenTIT2010}, it is decomposed according to the subvectors of $\bpsi$ into
\begin{equation} \label{eq:FIM_decomp} 
 \FIM_\bpsi = \left[ 
 \begin{array}{lll}
  \bLA & \bLB\R &\bLB\I \\
  (\bLB\R)\T & \bLC' & \bO \\
  (\bLB\I)\T & \bO & \bLC' 
 \end{array}
 \right] = \left[
 \begin{array}{ll}
 \bLA & \bLB \\
 \bLB\T & \bLC
 \end{array}
 \right] .
\end{equation}
Its elements are defined as \cite{Kay1993}, for example (see also (\ref{eq:FIM_LA})),
\begin{equation*}
  [\bLB\R]_{k,k'}  = 
\E{\br|\bpsi}{ - \frac{\partial^2 \ln f(\br|\bpsi)}
{\partial \tau_k \partial \Re\alpha_{k'}}} 
\end{equation*}
which yields with (\ref{eq:sampledLHF})
\begin{align}
 [\bLA]_{k,k'}
& = 2 \Re \left\{\alpha_k \alpha^*_{k'}
\left(\frac{\partial \bs_{\tau_{k'}}}{\partial \tau_{k'}}\right)\Herm
\bC_\bn^{-1} \frac{\partial \bs_{\tau_k}}{\partial \tau_k}\right\} \label{eq:FIM_channel_A} \\
 [\bLB\R]_{k,k'} 
& = 2 \Re \left\{\alpha_k \big(\bs_{\tau_{k'}}\big)\Herm
\bC_\bn^{-1} \frac{\partial \bs_{\tau_k}}{\partial \tau_k}\right\} \\
 [\bLB\I]_{k,k'} 
& = 2 \Im \left\{\alpha_k \big(\bs_{\tau_{k'}}\big)\Herm
\bC_\bn^{-1} \frac{\partial \bs_{\tau_k}}{\partial \tau_k}\right\} \\
 [\bLC']_{k,k'}
& = 2 \Re\left\{ \big(\bs_{\tau_{k}}\big)\Herm
\bC_\bn^{-1} \bs_{\tau_{k'}}\right\}.
\end{align}

These equations can be used to numerically evaluate the FIM without further assumptions. The CRLB can thus be evaluated,
but the inverse of the covariance matrix $\bCn$, which is needed as a whitening operator \cite{VanTrees1968} to
account for the non-stationary DM process, limits the insight it can possibly provide. More insight can be gained under
the assumption that the received deterministic MPCs $\{\alpha_k s(t - \tau_k)\}$ are orthogonal, which occurs in
practice when MPCs are non-overlapping.

\subsubsection{Orthogonal MPCs}\label{sec:LHF_orthoMPCs}

In this case, the columns of the signal matrix $\bS$ are orthogonal and $\bLA$ becomes  diagonal (since $\bC_\bn^{-1}$
is symmetric). Furthermore, $[\bLB]_{k,k'}$ is zero (due to the symmetry of the autocorrelation function of $s(t)$) and
as a consequence $[\bLC]_{k,k'}$ is not needed. The elements of $\bLA$ can then be written as (see 
Appendix~\ref{app:lhf})
\begin{align}\label{eq:FIM_channel_A_ortho}
  [\bLA]_{\mpcidxsym,\mpcidxsym} = 8\pi^2\beta^2 \SINR\mpcidx \gamma\mpcidx
\end{align}
where $\beta^2 = \int_f f^2 |S(f)|^2 \mathrm{d}f$ is the effective (mean square) bandwidth of the energy-normalized
transmit pulse $s(t) \stackrel{\mathcal{F}}{\longleftrightarrow} S(f)$,
\begin{equation}\label{eq:snr}
  \SINR\mpcidx :=  \frac{\big|\alpha\mpcidx\big|^2}{N_0+
  \Tp S_\nu(\tau\mpcidx ) }
\end{equation}
is the signal-to-interference-plus-noise ratio (SINR) of the $\mpcidxsym$-th MPC, and 
$\gamma\mpcidx$ is the so-called bandwidth extension factor. The product of these three factors quantifies the delay
information provided by the $k$-th MPC. It hence provides the following insight for the investigated estimation
problem:
The interference term $\Tp S_\nu(\tau\mpcidx )$ is determined by the PDP of DM $S_\nu(\tau\mpcidx )$ at the delay
$\tau\mpcidx$ of the MPC. It scales with the effective pulse duration $\Tp$ of the pulse $s(t)$, the reciprocal of its
equivalent Nyquist bandwidth $B_\mathrm{N}=1/\Tp$. An increased bandwidth is hence beneficial to suppress DM.

The bandwidth extension quantifies the SINR-gain due to the whitening operation. It is defined as
$\gamma\mpcidx = \beta\mpcidx^2/\beta^2$, where $\beta\mpcidx^2$  is the mean square bandwidth of the whitened pulse,
\begin{equation}
\label{eq:beta_k}
\beta\mpcidx^2 = \int_f f^2|\mathrm{S}(f)|^2 \frac{N_0 + \Tp S_\nu(\tau\mpcidx)}{N_0 + |\mathrm{S}(f)|^2 S_\nu(\tau\mpcidx)}\mathrm{d}f .
\end{equation}
If the pulse has a block spectrum, we have  (due to the energy normalization of $s(t)$) $|S(f)|^2 = \Tp$ for $|f|
\le B_\mathrm{N}/2$, hence $\beta\mpcidx^2=\beta^2$ and $\gamma\mpcidx = 1$. I.e., in this case, there is \textit{no}
bandwidth extension due to whitening\footnote{This specialization was assumed in our previous paper
\cite{WitrisalICC2012}.}. The same holds if DM is negligible, i.e. $N_0 \gg \Tp S_\nu(\tau\mpcidx)$. For the asymptotic
case that AWGN is negligible, i.e. $|\mathrm{S}(f)|^2 S_\nu(\tau\mpcidx) \gg N_0$, we drop $N_0$ in (\ref{eq:beta_k}) and
get a block spectrum that corresponds to the \textit{absolute} bandwidth of $S(f)$.

In general, $\gamma\mpcidx$ is a function of the interference-to-noise ratio (INR) $\Tp S_\nu(\tau\mpcidx) / N_0$ and
can be evaluated numerically. Closed-form results can be given for special cases. E.g. for a root-raised-cosine pulse
with roll-off factor $\rolloff$, we have $\beta^2 = B_\mathrm{N}^2(\frac{1}{12} + \frac{\pi^2-8}{4\pi^2}\rolloff^2)$
which scales slightly with $\rolloff$. In the asymptotic case where DM dominates, we get $\beta\mpcidx^2=
\frac{(1+\rolloff)^3}{12} B_\mathrm{N}^2$. Hence the bandwidth extension due to the whitening operation can result in an
SINR gain of up to about 7~dB at $\rolloff=1$. Numerical evaluation shows a $\gamma_k$ of 4~dB at $\rolloff=0.6$ and INR
of 15~dB.

For further analysis, we define the extended SINR 
\begin{equation}\label{eq:extSINR}
  \widetilde{\SINR}\mpcidx = \SINR\mpcidx \gamma\mpcidx
\end{equation}
which quantifies the delay information provided by MPC $k$ as a function of the signal, interference, and noise levels. 

\subsection{Position Error Bound}

The FIM $\FIM_\bpsi$ of the signal model parameters quantifies the information gained from the measurement $\br$. The
position-related part of this information lies in the MPC delays $\btau$, which are a function of the position
$\pind{}$. To compute the position error bound (PEB), the square-root of the trace of the CRLB on the position error, we
need the upper left $2\times2$ submatrix of the inverse of FIM $\FIM_\btheta$,
\begin{equation}\label{eq:PEB}
  \mathcal{P}\{\bp\} = \sqrt{\tr\left\{ \left[\FIM_\btheta^{-1}\right]_{2\times 2}\right\}} = \sqrt{\tr\left\{
  \FIM_\bp^{-1}\right\}},
\end{equation}
which can be obtained with \eqref{eq:ParmTrans} and \eqref{eq:JacobianGenereal} using the blockwise inversion lemma.
This results in the so-called \emph{equivalent} FIM (EFIM) $\FIM_\bp$ \cite{ShenTIT2010}, 
\[
 \FIM_\bp = \bH\T \big(\bLA - \bLB \big(\bLC\big)^{-1} \bLB\T\big) \bH ,
\]
which represents the information relevant for the position error
bound. Matrix $\bH = \partial \btau / \partial \bp$ is the submatrix of Jacobian (\ref{eq:JacobianGenereal}) that
relates to the position-related information, the derivatives of the delay vector $\btau$ w.r.t. postition $\bp$. It
describes the variation of the signal parameters w.r.t. the position and can assume different, scenario-dependent forms,
depending on the roles of anchors and agents. General expressions for these \emph{spatial delay gradients} are derived
in the next section.

\section{Spatial Delay Gradients}
\label{sec:geometry}
The following notations are used to find the elements of matrix $\bH$: $\bp\agentidx \in \mathbb{R}^2$ is the
position of the $\agentidxsym$-th agent, where $\agentidxsym \in \mathcal{N}_\agentidxsym = \{1,2, \ldots, \agentidxub
\}$. $\bp_1\anchoridx \in \mathbb{R}^2$ is the position of the $\anchoridxsym$-th fixed anchor, $\anchoridxsym \in
\mathcal{N}_\anchoridxsym = \{\agentidxub+1, \ldots, \agentidxub+\anchoridxub\}$, with VAs at  positions
$\bp\mpcidx\anchoridx \in \mathbb{R}^2$. In the cooperative scenario, we replace $\anchoridxsym$ with an arbitrary index
$\arbsymb$ to cover fixed anchors as well as agents which act as anchors. The corresponding VAs are at $\bp\mpcidx\arb
\in \mathbb{R}^2$. To describe gradients w.r.t. anchor or agent position, we use an index $\arbsymbtwo$, introducing
$\bp\arbtwo\in \mathbb{R}^2$. 

The delay of the $\mpcidxsym$-th MPC is defined by the distance between the $\mpcidxsym$-th
VA and the $\agentidxsym$-th agent, 
\begin{align}
  \tau\mpcidx^{(\arbsymb,\agentidxsym)} 
  &= \frac{1}{c} \big\| \bp\agentidx - \bp\mpcidx\arb \big\| \\ 
  &= \frac{1}{c} \sqrt{\big(x\agentidx - x\mpcidx\arb \big)^2 + \big(y\agentidx - y\mpcidx\arb\big)^2} .
\end{align}
The angle of vector $( \bp\agentidx - \bp\mpcidx\arb )$ is written as $\phi\mpcidx^{(\arbsymb,\agentidxsym)}$.
To describe the relation between the signal parameter $\tau\mpcidx^{(\arbsymb,\agentidxsym)}$ and the geometry, we need
to analyze the spatial delay gradient, the derivative of the delay $\tau\mpcidx^{(\arbsymb,\agentidxsym)}$ w.r.t. 
position $\bp\arbtwo$, 
\begin{align}\label{eq:gradintermediate}
  \bh\mpcidx^{(\arbsymb,\arbsymbtwo, \agentidxsym)} = & \fp{\tau\mpcidx^{(\arbsymb,\agentidxsym)}}{\bp\arbtwo} 
= \f{1}{c} \fp{\big\| \bp\agentidx - \bp\mpcidx\arb \big\|}{\bp\arbtwo} 
\nonumber \\
   \ =&\f{1}{c}  \fp{\big(x\agentidx - x\mpcidx\arb\big)}{\bp\arbtwo} 
\f{x\agentidx - x\mpcidx\arb}{\big\| \bp\agentidx - \bp\mpcidx\arb
    \big\|} \  \nonumber \\
   \ \hphantom{=} \ \ &+\f{1}{c} \fp{\big(y\agentidx - y\mpcidx\arb \big)}{\bp\arbtwo}  \f{y\agentidx -
y\mpcidx\arb}{\big\| \bp\agentidx - \bp\mpcidx\arb   \big\|}  \nonumber \\
  \ = & \f{1}{c} \Big( \delta_{m,\arbsymbtwo} \bI_2 - \delta_{\arbsymbtwo,\arbsymb}
  \fp{\bp\mpcidx\arb}{\bp\arb}\Big)\T  \be\Big(\phi\mpcidx^{(\arbsymb,\agentidxsym)}\Big) 
\end{align}
where $\be(\phi) := [\cos(\phi),\sin(\phi)]\T$ is a unit vector in direction of the argument angle and
$\delta_{m,\arbsymbtwo}$ is the Kronecker delta.Using \eqref{eq:vatxjaco} for the Jacobian ${\bp\mpcidx\arb}/{\bp\arb}$
of a VA position w.r.t. its respective anchor's position from Appendix \ref{app:geometry}, we get
\begin{align}\label{eq:gradresult}
  &\bh\mpcidx^{(\arbsymb,\arbsymbtwo, \agentidxsym)} = \\
  &\qquad \f{1}{c} \Big[\delta_{m,\arbsymbtwo}\be \Big(\phi\mpcidx^{(\arbsymb,\agentidxsym)}\Big) \ -
  \delta_{\arbsymbtwo,\arbsymb}  \be \Big((-1)^{Q\mpcidx\arb}
  \phi\mpcidx^{(\arbsymb,\agentidxsym)} + 2\eff\mpcidx\arb \Big) \Big], \nonumber
\end{align}
where the first summand represents the influence of the agent position while the second summand is linked to the anchor
position. \Erik{ The parameter $\eff\mpcidx\arb$ (see Appendix \ref{app:geometry}) describes
the effective wall angle of the $\mpcidxsym$-th MPC w.r.t. to the $\arbsymbtwo$-th anchor (or agent) and
$Q\mpcidx\arb$ represents the according VA order.}{} We stack the transposed gradient vectors (\ref{eq:gradresult}) for
the entire set of multipath components in the gradient matrix 
$  \bH^{(\arbsymb,\arbsymbtwo, \agentidxsym)} \in \mathbb{R}^{\mpcidxub^{(\arbsymb,\agentidxsym)} \times 2}$ and the matrices for
all the agents' derivatives into matrix $\bH^{(\arbsymb, \agentidxsym)} \in
\mathbb{R}^{\mpcidxub^{(\arbsymb,\agentidxsym)} 
\times2\agentidxub}$. 

The following specializations will be used:
\subsubsection{Bistatic scenario} \label{sec:geometry_bistatic} $\mpcidxsym = 1,\ldots,\mpcidxub^{(\arbsymb, \agentidxsym)}$

\paragraph{The gradient with respect to the agent} \label{sec:geometry_agent} This case describes the derivatives of delay $\tau^{(\arbsymb, \agentidxsym)}$ w.r.t. the agent position, i.e. $\arbsymbtwo = \agentidxsym$, yielding the gradient
\begin{align}\label{eq:bistaticfixedanchor}
  \hspace{-.5cm}
  \bh\mpcidx^{(\arbsymb,\agentidxsym,\agentidxsym)} =\fp {\tau^{(\arbsymb, \agentidxsym)}}{\bp\agentidx}
  =\f{1}{c}\be \Big(\phi\mpcidx^{(\arbsymb, \agentidxsym)} \Big)
\end{align}
which represents a vector pointing from an agent to the $\mpcidxsym$-th VA of the according anchor. We
define the gradient matrix $\GT^{(\arbsymb,\agentidxsym)} =\bH^{(\arbsymb,\agentidxsym,\agentidxsym)} \in
\mathbb{R}^{\mpcidxub^{(\arbsymb, \agentidxsym)} \times 2}$. 
\paragraph{The gradient with respect to the anchor} \label{sec:geometry_anchor} In this case, the derivatives w.r.t. the
anchor position $\bp^{(\arbsymb)} = \bp_1^{(\arbsymb)}$ are  described, i.e. $\arbsymbtwo = \arbsymb$. For the
$\mpcidxsym$-th MPC, the gradient is expressed as
\begin{align}\label{eq:bistaticmovedanchor}
  \hspace{-.5cm}
  \bh\mpcidx^{(\arbsymb,\arbsymb,\agentidxsym)} &= \fp {\tau\mpcidx^{(\arbsymb, \agentidxsym)}}{\bp^{(\arbsymb)}}
  \\
  & = -\f{1}{c}\be \Big((-1)^{Q\mpcidx^{(\arbsymb)}}
  \phi\mpcidx^{(\arbsymb, \agentidxsym)} + 2\eff\mpcidx^{(\arbsymb)} \Big)
    = \f{1}{c}\be \Big(\phi\mpcidx^{(\agentidxsym,\arbsymb)} \Big) \nonumber
\end{align}
which in this case is a vector pointing from an agent acting as anchor to the $\mpcidxsym$-th
VA of a cooperating agent. The proof for the final equality can be obtained graphically. The gradient
matrix is $\GR^{(\arbsymb,\agentidxsym)}= \GT^{(\agentidxsym,\arbsymb)} = \bH^{(\arbsymb,\arbsymb,\agentidxsym)} \in
\mathbb{R}^{\mpcidxub^{(\arbsymb, \agentidxsym)} \times 2}$.
\subsubsection{Monostatic scenario}\label{sec:geometry_mono} Here we restrict the VA set to $\mpcidxsym =
2,\ldots,\mpcidxub\agentagentidx$, the agent is as well the anchor, $\arbsymb = \agentidxsym$, and
both move synchronously, $\arbsymbtwo = \agentidxsym$, i.e., the two terms in (\ref{eq:gradresult}) interact with
each other. The gradient  
\begin{align}\hspace{-5cm} \label{eq:monograd}   
  &\bh\mpcidx^{(\agentidxsym,\agentidxsym,\agentidxsym)} = \fp{\tau\mpcidx\agentagentidx}
  {\bp\agentidx}  \\
  &= \f{1}{c} \left( \be \Big( \phi\mpcidx\agentagentidx \Big)-
  \be \Big((-1)^{Q\mpcidx\agentidx}
  \phi\mpcidx\agentagentidx + 2\eff\mpcidx\agentidx \Big) \right) \nonumber \\
  & = \left\{\!\begin{array}{ll}
      \f{2}{c}  \sin\! \Big(\eff\mpcidx\agentidx \Big)
      \be \Big(\phi\mpcidx\agentagentidx + \eff\mpcidx\agentidx - \f{\pi}{2} \Big) 
	&  \ \text{If} \ Q\mpcidx\agentidx \text{is even} \\
      \f{2}{c}  \sin\! \Big(\eff\mpcidx\agentidx -
      \phi\mpcidx\agentagentidx\Big)
      \be \Big(\eff\mpcidx\agentidx  -  \f{\pi}{2} \Big) 
	&  \ \text{If} \ Q\mpcidx\agentidx \text{is odd} \nonumber
      \end{array} \right.
\end{align}
has been decomposed---as shown in  Appendix \ref{app:mono}---into a magnitude term $0 \leq\big\|
\bh\mpcidx^{(\agentidxsym,\agentidxsym,\agentidxsym)}\big\|  \leq \f{2}{c}$ and a resulting direction vector. Both
depend on the angle $\phi\mpcidx\agentagentidx$, the VA order, and the angles of all contributing walls comprised
in $\eff\mpcidx\agentidx$. The gradient matrix is $\GM\agentidx=\bH^{(\agentidxsym,\agentidxsym,\agentidxsym)} \in
\mathbb{R}^{(\mpcidxub\agentagentidx-1)\times2}$. 
 
\emph{The following interpretations apply for the monostatic case:} Single reflections ($Q\mpcidx\agentidx = 1$,
$\eff\mpcidx\agentidx = \phi\mpcidx\agentagentidx \pm \f{\pi}{2}$) and reflections on rectangular corners
($Q\mpcidx\agentidx = 2$, $\eff\mpcidx\agentidx = \pm \f{\pi}{2}$) constitute important types of monostatic VAs. Both
have $\pa\tau\mpcidx\agentagentidx/\pa\bp\agentidx = \f{2}{c}\be(\phi\mpcidx\agentagentidx)$, which is twice as much
spatial sensitivity of delays as in the bistatic cases \eqref{eq:bistaticfixedanchor} and
\eqref{eq:bistaticmovedanchor}, thus providing higher ranging information. The simplest case of a vanishing gradient
(magnitude zero) is a second-order reflection between parallel walls ($Q\mpcidx\agentidx = 2$, $\eff\mpcidx\agentidx =
0$).

\section{CRLB on the Position Error}
\label{sec:CRLB_PE}
In this Section, the CRLB on the position error is derived for the three scenarios Multipath-Sync, Multipath-NSync, and
a Multipath-Coop scenario.

Using a stack vector $\bPsi =[\bT\T,\Re{\bA}\T,\Im{\bA}\T]\T$ of the signal parameters for all relevant nodes, 
with $\bT$ combining the delays and $\bA$ combining the amplitudes, the Jacobian \eqref{eq:JacobianGenereal} has the
following general structure. 
\begin{align}\label{eq:Jacobian}
  \JacMat &= \frac{\partial\bPsi}{\partial\bTheta}  
= \left[
  \begin{array}{ccc}
 \bH & \bL & \bZero \\
 \bZero & \bZero & \bI
  \end{array}\right]\\
  &= \left[ 
  \begin{array}{cccc}
    \pa \bT / \pa \bP & \pa \bT / \pa \clockoffvec & \pa \bT/ \pa \Re\bA & \pa \bT / \pa \Im\bA \\
    \pa \Re\bA / \pa \bP & \pa \Re\bA / \pa \clockoffvec & \pa \Re\bA / \pa \Re\bA & \pa \Re\bA/ \pa \Im\bA \\
    \pa \Im\bA/ \pa \bP & \pa \Im\bA / \pa \clockoffvec &  \pa \Im\bA  / \pa \Re\bA & \pa \Im\bA / \pa \Im\bA
  \end{array}\right] \nonumber
\end{align}
Vector $\bTheta = [\bP\T, \clockoffvec\T ,\Re{\bA}\T,\Im{\bA}\T]\T$, spatial delay gradient $\bH = \pa \bT / \pa \bP
$, and gradient $\bL = \pa \bT / \pa \clockoffvec$ are specifically defined for the different cases in the following
subsections.

\subsection{Derivation of the CRLB for Multipath-Sync}
\label{sec:CRLB_TOA}
Assuming that only one agent is present in Multipath-Sync and Multipath-NSync, we drop the agent index $\agentidxsym$
so that $\bP = \bp$, and define $\mathcal{N}_\anchoridxsym = \{1,2,\ldots,\anchoridxub \}$. We use the geometry for
the bistatic scenario, case (a) Section \ref{app:geometry}. The clock-offset $\clockoffvec$ is considered to be known
and zero. Using a suitable signaling scheme\footnote{E.g conventional multiple access schemes, like time-division-multiple-access (TDMA).}, measurements $\br\anchoridx$ from all $\anchoridxub$ anchors are independent. Hence, the log-likelihood function is defined as
\begin{equation}
  \ln f(\bR | \bPsi) = \sum_{\anchoridxsym \in \mathcal{N}_\anchoridxsym} \ln f\big(\br\anchoridx |
  \btau\anchoridx, \balpha\anchoridx \big),
\end{equation}
where $\bR = \big[ \big(\br\jnum{1}\big)\T, \ldots, \big(\br\jnum{\anchoridxub} \big)\T \big]\T$ combines all
measurements and $\btau\anchoridx$ and $\balpha\anchoridx$ are the delay and amplitude vectors respectively,
corresponding to measurement $\br\anchoridx$. The Jacobian $\JacMat$ has the following structure, 
%
%
\begin{align}\label{eq:JacobianTdoaSynced}
 \JacMat =  \left[ 
 \begin{array}{ccc}
    \bH\jnum{1} _{\mpcidxub\jnum{1}\times 2}  &   \\
    \vdots  &  \\
    \bH\jnum{\anchoridxub} _{\mpcidxub\jnum{\anchoridxub} \times 2}  & \\
& \bI _{ D_\bI \times  D_\bI} 
 \end{array}\right],
\end{align}
where zero-matrices in the off-diagonal blocks are skipped for clarity and $D_\bI = 2\sum_{\anchoridxsym =
1}^\anchoridxub \mpcidxub\anchoridx$. The subblocks $\bH\anchoridx = \GT^{(\anchoridxsym,1)}$ account for the geometry
as described in Section \ref{sec:geometry}. Due to the independence of the measurements $\br\anchoridx$, the EFIMs $\FIM_\bp\anchoridx$ from the $\anchoridxub$ different anchors are additive. Using Equation (\ref{eq:ParmTrans}), we can write the EFIM as 
\begin{align}\label{eq:efim}
  \FIM_\bp &= \\
  &\sum_{\anchoridxsym \in \mathcal{N}_\anchoridxsym} \big(\bH\anchoridx \big)\T \Big(\bLA\anchoridx -
  \bLB\anchoridx \big(\bLC\anchoridx \big)^{-1} \big(\bLB\anchoridx \big)\T \Big) \bH\anchoridx \notag
\end{align}
where $\bLA\anchoridx$, $\bLB\anchoridx$, and $\bLC\anchoridx$ are subblocks of $\FIM_\bpsi\anchoridx$ defined in
(\ref{eq:FIM_decomp}). Expression (\ref{eq:efim}) simplifies when we assume no path overlap (i.e. orthogonality)
between signals from different VAs. In this case, $\bLB = \bO$ and $\bLA$ will be diagonal, as discussed in Section
\ref{sec:LHF_orthoMPCs} and we can then write
\begin{align}\label{eq:efim1}
  \FIM_\bp &= \sum_{\anchoridxsym \in \mathcal{N}_\anchoridxsym}\big(\bH\anchoridx \big)\T \bLA\anchoridx \bH\anchoridx
  \nonumber \\
  &\approx \frac{8 \pi^2 \beta^2}{c^2} 
  \sum_{\anchoridxsym \in
\mathcal{N}_\anchoridxsym}\sum_{\mpcidxsym=1}^{\mpcidxub\anchoridx}\widetilde{\SINR}\mpcidx\anchoridx
\bD_\mathrm{r}(\phi\mpcidx\anchoridx)
\end{align}
where $\widetilde{\SINR}\mpcidx\anchoridx$ is the extended SINR (eq. \ref{eq:extSINR}) for the $\anchoridxsym$-th anchor
and
\begin{equation}
 \bD_\mathrm{r}(\phi\mpcidx\anchoridx) = \be(\phi\mpcidx\anchoridx) \be(\phi\mpcidx\anchoridx)\T
\end{equation}
is called ranging direction matrix (cf. \cite{ShenTIT2010}), a rank-one matrix with an eigenvector in direction of
$\phi\mpcidx\anchoridx$. 

Valuable insight is gained from (\ref{eq:efim1}) and (\ref{eq:snr}). In particular,
\begin{itemize}
 \item Each VA (i.e. each deterministic MPC) adds some positive term to the EFIM in direction of
  $\phi\mpcidx\anchoridx$ and hence reduces the PEB in direction of $\phi\mpcidx\anchoridx$. 
 \item The $\widetilde{\SINR}\mpcidx\anchoridx$ determines the magnitude of this contribution  as discussed in
Section~\ref{sec:LHF_orthoMPCs} (cf. ranging  intensity  information (RII) in \cite{ShenTIT2010}). It is limited by
diffuse multipath---an effect that reduces with increased bandwidth---and it can show a significant gain due to the
interference whitening if the interference-to-noise ratio is large.
 \item The effective bandwidth $\beta$ scales the EFIM. Any increase corresponds to a decreased PEB.
\end{itemize}

Discussion of \emph{path overlap} (cf. \cite{ShenTIT2010}): 
\begin{itemize}
 \item $\tau\mpcidx-\tau_{\mpcidxsym'} \ll \Tp$: In this case the MPCs can not be distinguished and the
  position-related information is entirely lost.
  \item $\tau\mpcidx-\tau_{\mpcidxsym'} \approx \Tp$: In this case the MPCs are  correlated, but the
position-related
  information can still partly be used. 
The discrete-time formulation of the CRLB based on (\ref{eq:sampledLHF}) can quantify this information
gain, in contrast to our previous, continuous formulation in \cite{WitrisalICC2012}.
 \item $\tau\mpcidx-\tau_{\mpcidxsym'} \gg \Tp$: If this holds, the MPCs are considered to be orthogonal and
(\ref{eq:efim1}) can be used if it holds for all $k \ne k'$. 
\end{itemize}
%

\subsection{Derivation of the CRLB for Multipath-NSync}
\label{sec:CRLB_TDOA}
Next we consider the same setup as before, but assume the clock offsets $\clockoffvec$ to be unknown parameters.
The differences between arrival times still provide position information in this case. When using multiple anchors, we
distinguish two different scenarios where either the clocks of all anchors are synchronized
among each other, or alternatively no synchronization is present at all. While this does not affect the signal parameter
FIM, we need to take it into account when performing the parameter transformation. Apart from the partial derivatives
$\bL = \pa \bT / \pa \clockoffvec$, the terms of the Jacobian are identical for Multipath-Sync and Multipath-NSync,
resulting in
\begin{align}
  \JacMat =  \left[ 
  \begin{array}{cccc}
    \bH_{\mpcidxub\jnum{1} \times  2}\jnum{1} & \bL\jnum{1}_{\mpcidxub\jnum{1} \times  D_{\clockoffvec}} &  \\
    \vdots & \vdots &   \\
    \bH_{\mpcidxub\jnum{\anchoridxub} \times 2}\jnum{\anchoridxub} & \bL\jnum{\anchoridxub}_{\mpcidxub\jnum{\anchoridxub} \times  D_{\clockoffvec}} & \\
     &  &  \bI_{D_{\bI} \times D_{\bI}} 
  \end{array}\right],
  \label{eq:TdoaJacobian}
\end{align}
where $\bL\anchoridx = \pa \btau\anchoridx / \pa \clockoffvec$ and  $D_{\clockoffvec}$ is the length of 
$\clockoffvec$.

\textit{Synchronized anchors}: 
When assuming $\clockoff\jnum{1} = \dots = \clockoff\jnum{\anchoridxub} = \clockoff$, the vector $\clockoffvec$ reduces
to $\clockoffvec = \clockoff$. The derivatives of the arrival times with respect to the clock offset are then given by
$\bL\anchoridx = \bl_{\text{syn}}\anchoridx = [1,\dots,1]\T $. Applying the parameter transformation and computing the
block inverse similarly as in (\ref{eq:efim}) leads to additivity of the $3\times 3$ EFIMs $\FIM_{\bp,
\clockoff}\anchoridx$ for the extended parameter vector $[\bp\T, \clockoff]\T$ (see Appendix~\ref{app:TDOA}). When
neglecting path overlap this expression simplifies to
\begin{equation}\label{eq:Tdoa3x3Efim}
  \FIM_{\bp, \clockoff} 
  = \sum_{\anchoridxsym \in \mathcal{N}_\anchoridxsym} \FIM_{\bp, \clockoff}\anchoridx
  = 8 \pi^2 \beta^2 \sum_{\anchoridxsym \in \mathcal{N}_\anchoridxsym}\sum_{\mpcidxsym=1}^{\mpcidxub\anchoridx}
  \widetilde{\SINR}\mpcidx\anchoridx \bD_{\mathrm{r},\clockoff}(\phi\mpcidx\anchoridx) ,
\end{equation}
with
\[
  \bD_{\mathrm{r},\clockoff}(\phi\mpcidx\anchoridx) = \bv \bv\T, \ \bv = \left[\f{1}{c} \cos(\phi\mpcidx\anchoridx), \
  \f{1}{c}\sin(\phi\mpcidx\anchoridx), \ 1 \right]\T .
\]
The inner sum in \eqref{eq:Tdoa3x3Efim} reveals that the $3 \times 3$ EFIMs $\FIM_{\bp, \clockoff}\anchoridx$ are in
canonical form. Since $\bD_{r,\clockoff}$ is a positive semidefinite matrix, it highlights that each VA adds
information for the estimation of $\bp$ and $\clockoff$, scaled by its extended $\widetilde{\SINR}\mpcidx$ and $\beta$.

The EFIM $\FIM_{\bp}$ can be computed from $\FIM_{\bp, \clockoff}$  by again applying the blockwise inversion lemma. When
neglecting path overlap, the expression for $\FIM_{\bp}$ becomes
\begin{equation}\label{eq:Tdoa2x2Efim}
  \FIM_{\bp} 
  = \frac{8 \pi^2 \beta^2}{c^2} 
  \left[
  \sum_{\anchoridxsym\in \mathcal{N}_\anchoridxsym}
  \sum_{\mpcidxsym=1}^{\mpcidxub\anchoridx} 
  \widetilde{\SINR}\mpcidx\anchoridx \bD_{\mathrm{r}}(\phi\mpcidx\anchoridx) - \clockoffinflmat
  \right] ,
\end{equation}
where $\clockoffinflmat$ accounts for the (negative) influence of the clock offset estimation with 
\begin{align*}
  \clockoffinflmat &= 
  \frac{1}{\sum_{\anchoridxsym \in \mathcal{N}_\anchoridxsym}\sum_{\mpcidxsym=1}^{\mpcidxub\anchoridx}
  \widetilde{\SINR}\mpcidx\anchoridx}\clockoffinfleigvec \clockoffinfleigvec\T ,	\\
  \clockoffinfleigvec &=  
  \sum_{\anchoridxsym \in \mathcal{N}_\anchoridxsym}
  \sum_{\mpcidxsym=1}^{\mpcidxub\anchoridx}\widetilde{\SINR}\mpcidx\anchoridx \be(\phi\mpcidx\anchoridx) .
\end{align*}
Note that Multipath-NSync can theoretically achieve equal performance as Multipath-Sync under the (rather unlikely)
condition $\clockoffinfleigvec = \bO$. Otherwise $\clockoffinflmat$ reduces the information, and thereby increases the
PEB. 

\textit{Asynchronous anchors}:
When having $\clockoff\jnum{i} \neq \clockoff\jnum{j}$, $ \forall i \neq j$, $i,j \in \mathcal{N}_\anchoridxsym$, we
stack all clock offsets in the vector $\clockoffvec = [\clockoff\jnum{1}, \dots, \clockoff\jnum{\anchoridxub}]\T$.
The derivatives of the arrival times with respect to the clock offsets are then given by a gradient matrix $\bL = \pa
\bT / \pa \clockoffvec $ of size $\sum_{\anchoridxsym \in \mathcal{N}_\anchoridxsym} K\anchoridx \times J$ which stacks
submatrices $\bL_{\text{asyn}}\anchoridx$ with  one nonzero column $[\bL_{\text{asyn}}\anchoridx]_{i,\anchoridxsym} =
1,\ i=1,\dots , K\anchoridx$. This leads to an additivity of the $2 \times 2$ EFIMs as shown in
Appendix~\ref{app:TDOA}, i.e. $\FIM_\bp = \sum_{\anchoridxsym \in \mathcal{N}_\anchoridxsym}\FIM_\bp\anchoridx$. When
neglecting path overlap, $\FIM_\bp$ takes the form of \eqref{eq:Tdoa2x2Efim}, but with 
\begin{equation}\label{eq:Tdoa2x2Efimasync}
  \clockoffinflmat = \sum_{\anchoridxsym \in \mathcal{N}_\anchoridxsym} 
  \frac{1}{\sum_{\mpcidxsym=1}^{\mpcidxub\anchoridx}\widetilde{\SINR}\mpcidx\anchoridx}
  \clockoffinfleigvec \anchoridx \big( \clockoffinfleigvec\anchoridx \big)\T \ ,
\end{equation}
\[
  \clockoffinfleigvec\anchoridx = \sum_{\mpcidxsym=1}^{\mpcidxub\anchoridx}\widetilde{\SINR}\mpcidx\anchoridx
  \be(\phi\mpcidx\anchoridx).
\]
Again, equality with Multipath-Sync is obtained if each $\clockoffinfleigvec\anchoridx = \bO$, otherwise
the PEB is increased.

\subsection{Derivation of the CRLB for Multipath-Coop}
\label{sec:CRLB_Coop}
We assume $\agentidxub$ agents $\agentidxsym \in \mathcal{N}_\agentidxsym = \{1,2, \ldots, \agentidxub \}$ and
$\anchoridxub$ fixed anchors $\anchoridxsym \in \mathcal{N}_\anchoridxsym = \{\agentidxub+1,
\ldots,\agentidxub+\anchoridxub\}$, which cooperate with
one another. As outlined in the Introduction, every agent conducts a monostatic measurement, meaning it emits a pulse 
and receives the multipath signal reflected by the environment, and conventional bistatic measurements with
all other agents and the fixed anchors. 
All measurements are distributed such that every agent is able to exploit information from any of its received
and/or transmitted signals. The clock-offsets $\clockoffvec$ are considered to be zero. 

The signal parameter vectors for the $(\anchoridxsym,\agentidxsym)$-th received signal $\br\anchoragentidx$ are defined
as 
$  \btau\anchoragentidx =
  \big[\tau_1\anchoragentidx , \ldots, \tau_{\mpcidxub\anchoragentidx}\anchoragentidx 
  \big]\T
$
and
$  \balpha\anchoragentidx=\big[
  \alpha_1\anchoragentidx , \ldots, \alpha_{\mpcidxub\anchoragentidx}\anchoragentidx 
  \big]\T
$. For deriving the cooperative EFIM, we stack positions $\bp\agentidx$ of the $\agentidxub$ agents into the vector
\begin{equation}
  \bP = \big[\big(\bp\jnum{1}\big)\T, \ldots, \big(\bp\jnum{\agentidxub}\big)\T \big]\T \in
  \mathbb{R}^{2\agentidxub \times 1}
\end{equation}
and all measurements $\br\anchoragentidx$ in the vector
\begin{multline}
  \bR = \big[ \big(\br\jmnum{1}{1}\big)\T, \ldots, \big(\br\jmnum{1}{\agentidxub}\big)\T, \ldots, 
  \big(\br\jmnum{\agentidxub}{\agentidxub}\big)\T, \\
  \big(\br\jmnum{\agentidxub+1}{1}\big)\T,\ldots,\big(\br\jmnum{\agentidxub+\anchoridxub}
  {\agentidxub}\big)\T \big]\T \in \mathbb{C}^{D_\bR \times 1},
\end{multline}
where $D_\bR = \timestepsymub\agentidxub(\agentidxub+\anchoridxub)$. Further, we stack the 
signal parameters correspondingly in the vectors
\begin{multline}
  \hspace{-.2cm}
  \bT = \big[ \big(\btau\jmnum{1}{1}\big)\T, \ldots, \big(\btau\jmnum{1}{\agentidxub}\big)\T, \ldots,
\big(\btau\jmnum{\agentidxub + \anchoridxub}
  {\agentidxub}\big)\T \big]\T
\end{multline}
and
\begin{multline}
  \hspace{-.2cm}
  \bA = \big[ \big(\balpha\jmnum{1}{1}\big)\T, \ldots, \big(\balpha\jmnum{1}{\agentidxub}\big)\T, \ldots,
  \big(\balpha\jmnum{\agentidxub+\anchoridxub}{\agentidxub}\big)\T \big]\T
\end{multline}
of length $D_\bT = D_\bA = \sum_{\anchoridxsym \in ( \mathcal{N}_\agentidxsym \cup  \mathcal{N}_\anchoridxsym)}
\sum_{\agentidxsym \in
\mathcal{N}_\agentidxsym} \mpcidxub\anchoragentidx $ to construct vector $\bPsi = [\bT\T,\Re\bA\T,\Im\bA\T]\T$ .  The
corresponding joint log-likelihood function, assuming independent measurements $\br\anchoragentidx$ between the
cooperating nodes, is defined as
\begin{equation}\label{eq:LHF_coop}
  \ln f(\bR | \bPsi) = \sum_{\anchoridxsym \in (\mathcal{N}_\agentidxsym\cup  \mathcal{N}_\anchoridxsym)}
\sum_{\agentidxsym \in
  \mathcal{N}_\agentidxsym} \ln f\big(\br\anchoragentidx | \btau\anchoragentidx, \balpha\anchoragentidx\big) . 
\end{equation}
The EFIM $\FIM_\bP$ is described by (see Appendix \ref{app:coop})
\begin{equation}\label{eq:FIM_coop}
  \FIM_\bP =\sum_{\anchoridxsym \in( \mathcal{N}_\agentidxsym\cup  \mathcal{N}_\anchoridxsym)}
  \sum_{\agentidxsym \in \mathcal{N}_\agentidxsym}\big(\bH\anchoragentidx \big)\T \bLambda\anchoragentidx
  \bH\anchoragentidx
\end{equation}
where
\begin{equation}\label{eq:efim_coop}
  \bLambda\anchoragentidx = \bLA\anchoragentidx - \bLB\anchoragentidx \big(\bLC\anchoragentidx \big)^{-1}
  \big(\bLB\anchoragentidx \big)\T
\end{equation}
yields the sub-blocks $\FIM_\bpsi\anchoragentidx$ of the FIM  for the likelihood function
(\ref{eq:LHF_coop}), for independent  measurements, and $\bH\anchoragentidx$ are the spatial delay
gradients\footnote{\Erik{Multipath-Coop can be seen as the most general setup, if clock offset issues are also included.
This can be done by combining the results of Multipath-NSync and Multipath-Coop by replacing $\bH\anchoragentidx$ with
$\bG\anchoragentidx = [\bH\anchoragentidx,\bL\anchoragentidx]$ (see Appendix~\ref{app:TDOA}), which accounts for the
geometry and clock offset. For monostatic measurements $\bL^{\agentidxsym,\agentidxsym} = \mathbf{0}$. }{}} of the
Jacobian
\begin{align}\label{eq:Jacobian_coop}
  \JacMat \!=\!  \left[\! 
  \begin{array}{cccc}
    \bH_{\mpcidxub\jmnum{1}{1} \times 2\agentidxub}\jmnum{1}{1} &  \\
    \vdots & \\
    \bH_{\mpcidxub\jmnum{1}{\agentidxub} \times 2\agentidxub}\jmnum{1}{\agentidxub} & \\
    \vdots & \\
    \bH_{\mpcidxub\jmnum{\agentidxub+\anchoridxub}{\agentidxub} \times
    2\agentidxub}\jmnum{\agentidxub+\anchoridxub}{\agentidxub} & \\
    & \bI_{D_\bI \times D_\bI} 
  \end{array}\!\right],
\end{align}
where $D_\bI = 2D_\bA $.\footnote{Assuming no path overlap, (\ref{eq:FIM_coop}) can be
simplified as in (\ref{eq:efim1}), using the result from Appendix~\ref{app:lhf}.}
As shown in Appendix \ref{app:coop}, one gets the following final result for the EFIM $\FIM_\bp$ for all agents
\begin{align}\label{eq:EFIM_coop}
  &\FIM_\bP = 
\left[ \begin{array}{cccr}
  \FIMMo\jnum{1} \!+\! 2\FIMAg \jnum{1} \!+\! \FIMAn \jnum{1} &
  2\FIMC \jmnum{1}{2} & \ldots & 2\FIMC \jmnum{1}{\agentidxub} \\
  2\FIMC \jmnum{2}{1} &  \multicolumn{2}{l} \ddots & \\
  \!\vdots\! &&& \\
  2\FIMC \jmnum{\agentidxub}{1} & \multicolumn{3}{r}{\FIMMo\jnum{\agentidxub} \!+\! 2\FIMAg \jnum{\agentidxub} \!+\!
  \FIMAn \jnum{\agentidxub}}
  \end{array} \right] .
\end{align}
The diagonal blocks $\FIMAg \jnum{\arbsymbtwo} = \sum_{\agentidxsym \in
\mathcal{N}_\agentidxsym \backslash \{\arbsymbtwo\}}\big(\GT^{(\agentidxsym,\arbsymbtwo)}\big)\T
\bLambda^{(\agentidxsym,\arbsymbtwo)} \GT^{(\agentidxsym,\arbsymbtwo)}$ account for the bistatic measurements between
agent $\arbsymbtwo$ and all other agents, $\FIMAn \jnum{\arbsymbtwo} =  \sum_{\anchoridxsym
\in \mathcal{N}_\anchoridxsym} \big(\GT^{(\anchoridxsym,\arbsymbtwo)}\big)\T \bLambda^{(\anchoridxsym,\arbsymbtwo)}
\GT^{(\anchoridxsym,\arbsymbtwo)}$ account for the bistatic measurements between agent $\arbsymbtwo$ and all fixed
anchors, and $\FIMMo \jnum{\arbsymbtwo} =\big( \GM^{(\arbsymbtwo)}\big)\T\bLambda^{(\arbsymbtwo,\arbsymbtwo)}
\GM^{(\arbsymbtwo)}$ account for the monostatic measurement of agent $\arbsymbtwo$. The off-diagonal blocks $\FIMC
\jmnum{\arbsymbtwo}{\arbsymbtwo'} =\big(\GT^{(\arbsymbtwo',\arbsymbtwo)}\big)\T\bLambda^{(\arbsymbtwo',\arbsymbtwo)}
\GT^{(\arbsymbtwo,\arbsymbtwo')}$ account for the uncertainty about the cooperating agents in their role as anchors 
(cf. (\ref{eq:EFIM_offdiag}) and (\ref{eq:EFIM_ondiag})). This has a negative effect on the localization performance of
the agents. The factors of two in (\ref{eq:EFIM_coop}), related to the EFIM of measurements inbetween agents, results
from the fact that those measurements are performed twice. This simplifies the notations in this section. If such
repeated measurements are avoided, the same result would apply but with these factors removed. 

Finally, the CRLB on position $\bp^{(\arbsymbtwo)}$ of agent $\arbsymbtwo$ is 
\begin{equation}\label{eq:PEB_Coop}
  \mathcal{P}\{ \bp^{(\arbsymbtwo)}\} = \sqrt{\tr\left\{ \left[\FIM_\bP^{-1}\right]_{2\times
  2}\jmnum{\arbsymbtwo}{\arbsymbtwo}\right\}}.
\end{equation}

\section{Results}
\label{sec:results}
\begin{figure}[t!]
  \centering
  \includegraphics[width=1\columnwidth]{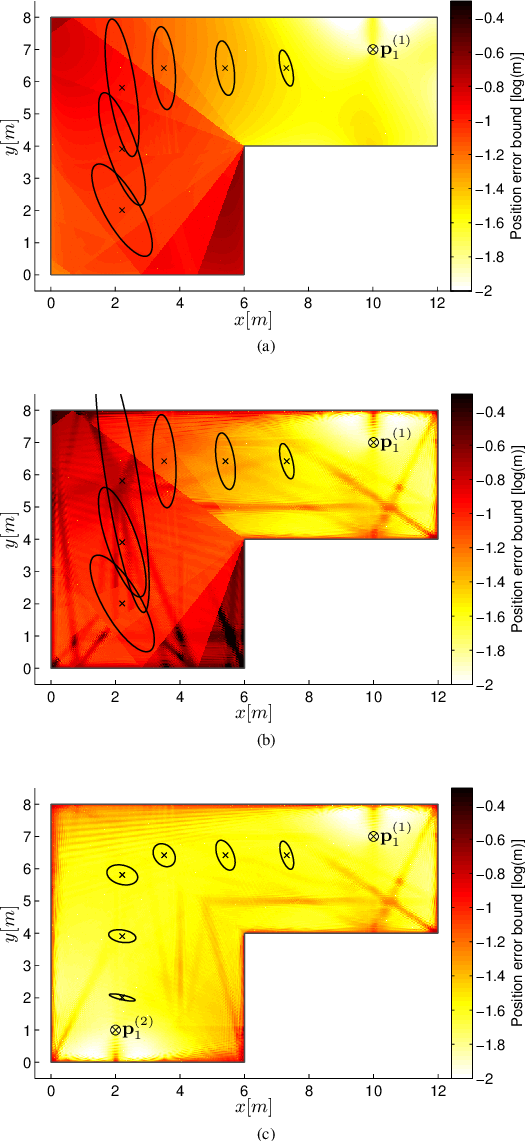}
  \caption{Logarithmic PEB  \eqref{eq:PEB} for Multipath-Sync with \vu{\Tp=1}{ns} over the example room for VAs
  up to order two. (a) One anchor at $\pind{1}^{(1)}$; path overlap neglected. (b) same as (a) but considering the
  influence of path overlap. (c) a second anchor has been introduced at $\pind{1}^{(2)}$; path overlap included. At some
  sample points, 20-fold standard deviation ellipses are shown. }
  \label{fig:peb2DToA}
\end{figure}

\graphicspath{{./figures/final_fig/}}
\begin{table}[t]
  \centering
  \caption{Channel parameters for numerical evaluations.} 
  \label{table:channelparameters}
  \begin{tabular}{ l|c|c|c|l}
    & Param. & \multicolumn{2}{c|}{Value for Room} & Description \\ 
    & & Valid. & Synth. & \\
    \hline \hline
    \multirow{2}{*}{\pbox{2cm}{Deterministic\\ MPCs}} & & \multicolumn{2}{c|}{2} & max. VA order \\ 
    & & \multicolumn{2}{c|}{\vu{3}{dB}} & attenuation per\\
  & & \multicolumn{2}{c|}{~} & reflection \\ \hline
    \multirow{4}{*}{\pbox{2cm}{Signal\\ parameters}} & $\fc$ & \vu{8}{GHz} & \vu{7}{GHz} & carrier freq. \\
    & $\Tp$ & \multicolumn{2}{c|}{\vu{1}{ns}, (\vu{0.5}{ns},\vu{2}{ns})} & pulse duration \\
    & & \multicolumn{2}{c|}{RRC} & pulse shape \\
    & $\rolloff$ & \multicolumn{2}{c|}{$0.6$} & roll-off factor \\ \hline
  %
    \multirow{4}{*}{\pbox{2cm}{PDP of diffuse \\ multipath}} & $\Omega_1$ & \vu{2.67\text{e}^{-6}}{}
&\vu{1.16\text{e}^{-6}}{} & norm. power \\
    & $\gamma_1$ &  \vu{10}{ns} & \vu{20}{ns} & \multirow{3}{*}{shape param.}\\
    & $\gamma_\text{rise}$ &  \vu{3}{ns} & \vu{5}{ns} &  \\
    & $\chi$ & \multicolumn{2}{c|}{$0.98$} &  \\ \hline
    $\mathrm{E}_\mathrm{LOS}/N_0$ &  & \multicolumn{2}{c|}{\vu{29.5}{dB} (at \vu{1}{m})} & LOS SNR
\\
   \hline
  \end{tabular}
\end{table}
\begin{table}
  \centering
  \caption{MPC SINRs for the validation environment, estimated from measured signals and computed from the
  channel model.}
  \label{table:SINR}
  \begin{tabular}{ p{3.2cm}|c|c|c}
         & \multicolumn{3}{c}{SINR (measurem.) / SINR (model) [dB] }  \\
     MPC & $\Tp=\vu{0.5}{ns}$ &  $\Tp=\vu{1}{ns}$ & $\Tp=\vu{2}{ns}$\\ 
    \hline \hline
    LOS Anchor 1                  & 23.1 / 25.8 & 24.7 / 24.7 & 23.2 / 23.7 \\
    lower wall                & 11.1 / 18.3 & 5.4 / 15.9  & 4.1 / 13.7 \\
    right window              & 13.5 / 12.6 & 7.6 / 10.2  & 6.9 / 7.7 \\
    upper wall               & 2.2 / 11.7  & -0.6 / 9.5  & 5.2 / 7.1 \\
    lower wall -- right win. & 9.5 / 7.3   & 7.6 / 4.9   & 4.9 / 2.4 \\
    \hline
    LOS Anchor 2                   & 25.9 / 26.4 & 26.0 / 25.3 & 26.5 / 24.2 \\
    right window             & 11.9 / 12.6 & 10.5 / 10.8 & 9.3 / 8.8 \\
    upper window             & 10.1 / 14.0 & 8.2 / 11.6  & 5.1 / 9.1 \\ 
    left wall                & 3.1 / 14.4  & 4.2 / 11.9  & 5.5 / 9.4 \\
    upper wall -- right win. & 10.6 / 5.7  & 11.7 / 3.9  & 3.5 / 1.8 \\
    upper win. -- left wall  & 7.2 / 9.7   & 4.8 / 7.3   & 2.1 / 4.8 \\
    \hline
  \end{tabular}
\end{table}
Computational results are presented in this section for two environments. We first validate the theoretical
results using experimental data for a room illustrated in Fig.~\ref{fig:peb2DIPIN} and then discuss in detail the
trade-offs of different measurement scenarios for a synthetic room shown in Fig.~\ref{fig:peb2DToA}.

For the transmit signal $s(t)$, we use a root-raised-cosine (RRC) pulse with unit energy and a roll-off factor $\rolloff
= 0.6$, modulated on a carrier at $\fc=\vu{7}{GHz}$ and $\fc=\vu{8}{GHz}$ (see Table \ref{table:channelparameters}). The computations are done for pulse durations of \vu{\Tp=0.5}{ns}, \vu{\Tp=1}{ns} and \vu{\Tp=2}{ns}. In the synthetic environments, we assume for all antennas   isotropic radiation patterns in the azimuth plane and gains of \vu{0}{dB}. The free-space pathloss has been modeled by the Friis equation. To account for the material impact, we assume \vu{3}{dB} attenuation per reflection. As in our previous paper
\cite{WitrisalICC2012}, the PDP of the DM is considered to be a fixed double-exponential function, as introduced by
\cite[eq. (9)]{KaredalTWC2007}. This choice reflects the common assumption of an exponential decay of the DM power and
also the fact that the LOS component is not impaired by DM as severely as MPCs arriving later \cite{SteinbockTAP2013}.
The model has been fitted in \cite{KaredalTWC2007}  to measurements collected in an industrial environment. We have used
$\chi = 0.98$ as in \cite{KaredalTWC2007} to describe the impact of DM on the LOS component and adapted
$\gamma_\text{rise}$ and $\gamma_1$ to reflect the smaller dimensions of our environments. Table
\ref{table:channelparameters} summarizes the parameters of the channel and signal models.

We would like to emphasize that this parametric model was introduced for simplicity and reproducibility, to
analyze the impact of DM on the PEB in various scenarios. In practice, the SINR values can be estimated from channel
measurements and used with the results from Section~\ref{sec:CRLB_PE} to compute the PEB for real environments. This
approach is used next to validate the theoretical results and the parametric channel model.

\subsection{Validation with Measurement Data: Multipath-Sync} 

\begin{figure}[t]
  \centering
  \includegraphics[width=1\columnwidth]{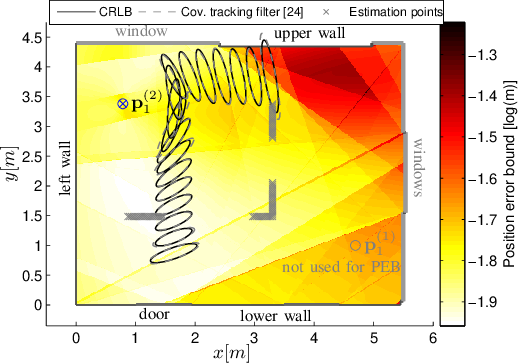}
  \caption{Logarithmic PEB \eqref{eq:PEB} for estimated SINRs in the validation environment using measured signals with
$\Tp=\vu{0.5}{ns}$ and $\fc=\vu{8}{GHz}$ and only MPCs corresponding to the anchor at $\pind{1}^{(2)}$. 30-fold standard
deviation ellipses are shown for the CRLB and a tracking algorithm (c.f. \cite{Meissner2014a}).}
  \label{fig:peb2DIPIN}
\end{figure}
The validation is conducted in an example environment shown in Fig.~\ref{fig:peb2DIPIN}, c.f. \cite{MeissnerICC2014}.
The MPC SINRs \eqref{eq:snr} are estimated from channel measurement data as discussed in
\cite{Meissner2014a,MeissnerEUCAP2012}, using fixed positions for two anchors and a set of ``estimation points'' for the
agent as illustrated in the figure. Table \ref{table:SINR} shows the obtained values for selected MPCs. It also lists
the corresponding SINRs computed from  the parametric channel model, with parameters given in
Table~\ref{table:channelparameters}. The choice of the parameters of the double exponential PDP of the DM has been made
to account for the smaller room dimensions in comparison to the synthetic environment used below.

The \textit{estimated SINRs} in Table \ref{table:SINR} show the relevance of the corresponding MPCs. The LOS is
the most significant one. Its SINR is approximately constant over all bandwidths used, indicating that it is only
slightly influenced by DM. The reflections at the windows and at the lower wall also provide significant
position-related information. A scaling with bandwidth---as suggested by \eqref{eq:snr}---is observable reasonably well.
Other MPCs provide less information, such as the left wall (plasterboard) and the upper wall. This is caused by a
reduced reflection coefficient, increased interference by DM, and increased variance of the  MPC amplitude over the
estimation points. Reference \cite{MeissnerPhD2014} contains further results supporting the presented findings based on
measurement data from other environments \cite{MeasureMINT2013}.

Table \ref{table:SINR} also shows that the \textit{parametric channel model} yields
realistic SINRs in many cases and therefore valid performance bounds. 
It has to be stressed that the global PDP model as used here cannot describe the 
local behavior of DM. However, based on the provided framework, it is straightforward to introduce more realism by
fitting separate parameterized or sampled models to any appropriate local area.

Figure \ref{fig:peb2DIPIN} shows the logarithmic PEB for the validation environment using the estimated SINRs from
Table~\ref{table:SINR} for Anchor~2 and $\Tp=\vu{0.5}{ns}$. Equation~\eqref{eq:efim1} has been employed to compute the
PEB, i.e. path overlap has been neglected and synchronization assumed. 
Clearly, one can observe from this figure the visibility regions and the relative importance (c.f. Table
\ref{table:SINR}) of specific MPCs. The PEB is better than 10~cm at almost the entire area. The  ellipses encode the
geometrically decomposed PEB with 30-fold standard deviation, computed from (\ref{eq:PEB}). Dashed ellipses are for a
multipath-assisted tracking algorithm \cite{Meissner2014a} that makes use of the estimated SINRs for properly weighting
the information from MPCs. It can be observed that both results match closely.

\subsection{Synthetic Environment}
\label{sec:mp_toa}

\begin{figure}[!ht]
  \centering
  \includegraphics[width=1\columnwidth]{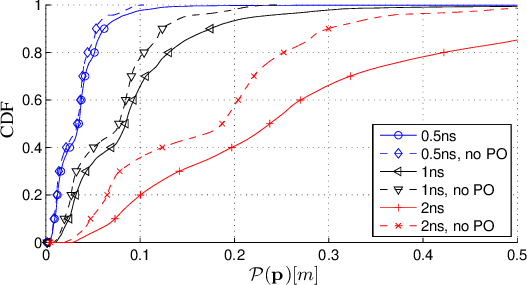}
  \caption{CDFs of the PEB  \eqref{eq:PEB} for Multipath-Sync, pulse durations \vu{\Tp = 0.5}{ns}, \vu{\Tp =
  1}{ns} and \vu{\Tp = 2}{ns}, and one anchor at $\bp_1\jnum{1}$. Path overlap is neglected in results marked by dashed
  lines.} 
  \label{fig:pebtoa1anchorVA2CDF}
\end{figure}
\begin{figure}[!ht]
  \centering
  \includegraphics[width=1\columnwidth]{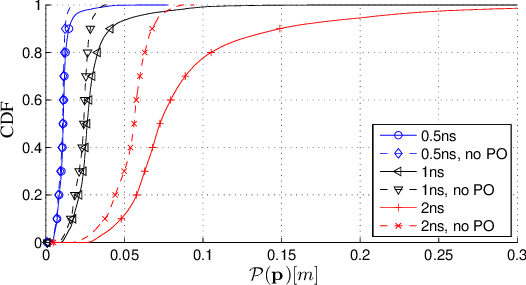}
  \caption{CDFs of the PEB  \eqref{eq:PEB} for Multipath-Sync, pulse durations \vu{\Tp = 0.5}{ns}, \vu{\Tp =
  1}{ns} and \vu{\Tp = 2}{ns}, and two anchors at $\bp_1\jnum{1}$ and $\bp_1\jnum{2}$. Path overlap is neglected in
  results marked by dashed lines.} 
  \label{fig:pebtoa2anchorVA2CDF}
\end{figure}
The synthetic environment shown in Fig.~\ref{fig:peb2DToA} is used to compare different measurement scenarios. The PEB
is evaluated across the entire room, assuming one or two fixed anchor at positions $\bp_1^{(1)} = [10,7]\T$ and
$\bp_1^{(2)} = [2, 1]\T$. We use a point grid with a resolution of \vu{2}{cm}, resulting in 180,000 points. VAs up to
order two are considered, unless otherwise specified.

\subsubsection{Multipath-Sync}
Fig.~\ref{fig:peb2DToA} shows the PEB over the floor-plan for Multipath-Sync and \vu{\Tp = 1}{ns}.
Figs.~\ref{fig:peb2DToA}(a) and (b) compare the simplified PEB neglecting path-overlap (cf. \eqref{eq:efim1}) with the
full PEB considering it (cf. \eqref{eq:efim}). A single anchor is employed in both cases at position $\pind{1}^{(1)}$,
yielding a PEB below \vu{10}{cm} for most of the area. One can clearly see the visibility regions of different
VA-modeled MPCs encoded by the level of the PEB. A valid PEB is obtained over the entire room even though the anchor is
partly not visible from the agent positions. If path-overlap is considered (Fig.~\ref{fig:peb2DToA}(b)) in the
computation of the CRLB, the adverse effect of room symmetries
is observable, corresponding to regions where deterministic MPCs overlap. In case of \emph{unresolvable} path overlap,
i.e. the delay difference of two MPCs is less than the pulse duration $\tau\mpcidx -
\tau_{\mpcidxsym'} \ll \Tp$, the information of the  components is entirely lost  (see
Section~\ref{sec:CRLB_TOA}). The ellipses illustrate the geometrically decomposed PEB with 20-fold  standard-deviation. 

Fig.~\ref{fig:peb2DToA}(c) shows the PEB with path-overlap for the same parameters but for two anchors. The error
ellipses clearly indicate that the PEB is much smaller and the impact of path overlap has been reduced. 

A quantitative assessment of this scenarios is given in Figs.~\ref{fig:pebtoa1anchorVA2CDF} and
\ref{fig:pebtoa2anchorVA2CDF}, showing the CDFs of the PEB for different pulse durations (\vu{\Tp = 0.5}{ns}, \vu{\Tp =
1}{ns} and \vu{\Tp = 2}{ns}). One can observe that the PEB increases vastly w.r.t. this parameter. The ``no PO'' results
account for the proportional scaling of Fisher information with bandwidth
and additionally for the increased interference power due to DM, both of which are clearly seen in approximation
(\ref{eq:efim1}). The influence of path overlap, which is neglected by (\ref{eq:efim1}), magnifies this effect even
further because its occurrence becomes more probable. It almost diminishes---on the other hand---for the shortest pulse
\vu{\Tp =0.5}{ns}. Over all, the error magnitude scales by a factor of almost ten, while the bandwidth is scaled by a
factor of  four.   

Our work in \cite{LeitingerICC2014, MeissnerICC2014, Meissner2014a} shows algorithms based on the presented signal model
that can closely approach these bounds. I.e. cm-level accuracy is obtained for \vu{90}{\%} of the estimates.

\subsubsection{Multipath-NSync}
\label{sec:mp_tdoa}

\begin{figure}[t]
  \centering
  \includegraphics[width=1\columnwidth]{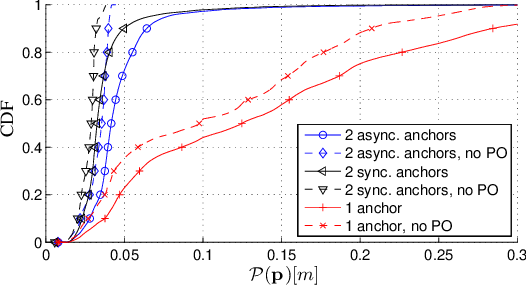}
  \caption{CDFs of the PEB \eqref{eq:PEB} for Multipath-NSync and different synchronization states at pulse
duration \vu{\Tp = 1}{ns}. Either two anchors are used at $\bp_1\jnum{1}$ and $\bp_1\jnum{2}$,
which can be synchronized or not, or just the first anchor.}
  \label{fig:pebtdoasync}
\end{figure}

Fig.~\ref{fig:pebtdoasync} compares the CDFs of the PEB for Multipath-NSync and different
synchronization states inbetween anchors, obtained from \eqref{eq:Tdoa2x2Efim}.  The CDFs are shown for either two
anchors at $\bp_1\jnum{1}$ and $\bp_1\jnum{2}$ which can be synchronized or not, or just the first anchor. A pulse
duration of \vu{\Tp = 1}{ns} is used. The performance deteriorates w.r.t. the Multipath-Sync case in
Figs.~\ref{fig:pebtoa1anchorVA2CDF} and \ref{fig:pebtoa2anchorVA2CDF}, which can be
explained by the fact that some of the delay information is used for clock-offset estimation, resulting in a loss of
\textit{position}-related information. A second anchor helps to counteract this effect. Here, one can recognize an
additional gain of information if the two anchors are synchronized. The impact of path overlap is smaller if
two anchors are used and even less pronounced if the anchors are synchronized.

\begin{figure}[!ht]
  \centering
  \includegraphics[width=1\columnwidth]{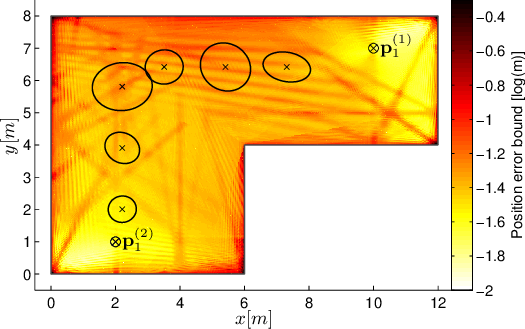}
  \caption{Logarithmic PEB \eqref{eq:PEB} for Multipath-NSync over the example room with \vu{\Tp=1}{ns}, using
  two asynchronous anchors at $\bp_1\jnum{1}$ and $\bp_1\jnum{2}$.  20-fold
  standard deviation ellipses are shown at some sample points.} 
  \label{fig:peb2Dtdoa}
\end{figure}

\begin{figure}[!ht]
  \centering
  \includegraphics[width=1\columnwidth]{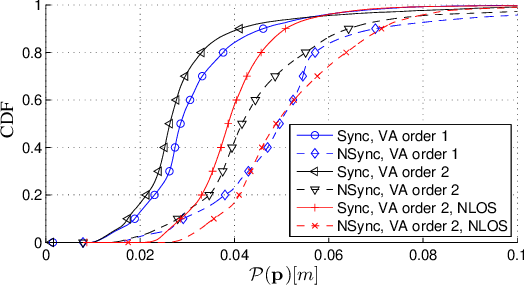}
  \caption{CDFs of the PEB in \eqref{eq:PEB} for Multipath-Sync and Multipath-NSync and \vu{\Tp =
1}{ns} with two anchors at $\bp_1\jnum{1}$ and $\bp_1\jnum{2}$. VAs of order one or two are considered; for the latter
case also for an artificial NLOS situation over the whole room.} 
  \label{fig:pebtdoavstoa}
\end{figure}

A qualitative representation of the PEB is shown in Fig.~\ref{fig:peb2Dtdoa} for
Multipath-NSync  over the example room, with two anchors at $\bp_1\jnum{1}$ and $\bp_1\jnum{2}$, and \vu{\Tp=1}{ns}.
Comparing this result with the synchronized case shown in Fig.~\ref{fig:peb2DToA}(c), one can observe an increase due to
the need of extracting syncronization information. Also, the impact of path overlap has increased.

Fig.~\ref{fig:pebtdoavstoa} compares Multipath-Sync and Multipath-NSync for the two-anchors case and \vu{\Tp=1}{ns},
considering VAs of order one or two and an NLOS scenario where the LOS component has been set to zero across the
entire room. One can observe the importance of the LOS component which usually has a significantly larger SINR and
provides thus more position-related information than MPCs arriving later. Increasing the VA order leads in general also
to an information gain. However, in a few cases this trend is reversed  since a larger VA-order can lead to more
positions with unresolvable path overlap. This occurs especially at locations close to walls and in corners.

\subsubsection{Multipath-Coop}
\label{sec:mp_coop}

\begin{figure}[ht]
  \centering
  \includegraphics[width=1\columnwidth]{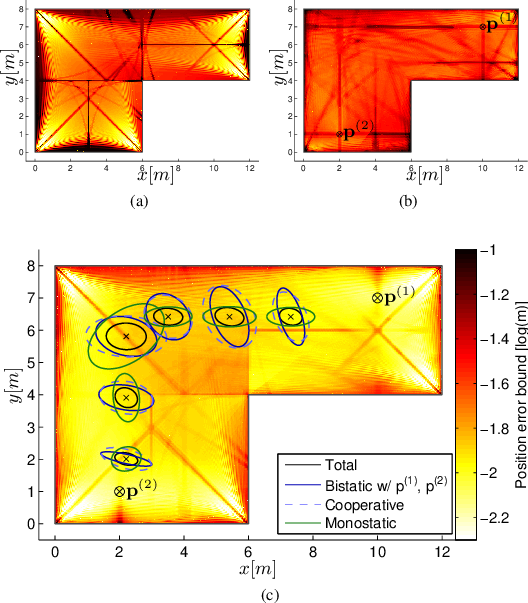}
  \caption{Logarithmic PEB  \eqref{eq:PEB_Coop} with \vu{\Tp=1}{ns} over the example room for three cooperating agents,
  two of which are resting at positions $\bp\jnum{1}$ and $\bp\jnum{2}$. The PEB is decomposed into its (a) monostatic
  and (b) cooperative components. Plot (c) shows the total PEB for Multipath-Coop. In (c), also the 40-fold standard
deviation ellipses are shown at some sample points for these three cases and---in addition---for the (bistatic) case 
with fixed anchors.}
\label{fig:peb2DCoop}
\end{figure}

\begin{figure}[ht]
  \centering
  \includegraphics[width=1\columnwidth]{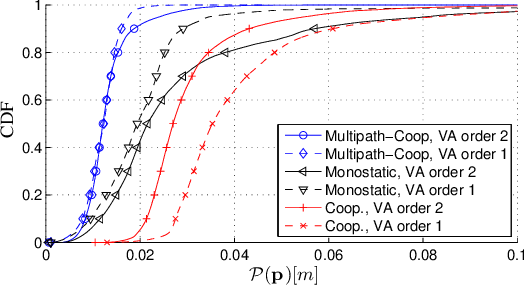}
  \caption{CDFs of the PEB \eqref{eq:PEB_Coop} for Multipath-Coop with \vu{\Tp=1}{ns}, for VAs of order one and two, analyzing contributions of different measurements.} 
  \label{fig:pebCoopParts}
\end{figure}

Fig.~\ref{fig:peb2DCoop} contains 2D-plots of the different contributions to the PEB in \eqref{eq:PEB_Coop} for the
cooperative case. The PEB has been evaluated for Agent 3  across the entire room with two  resting, cooperating agents
at $\bp\jnum{1}$ and $\bp\jnum{2}$. In Fig.~\ref{fig:peb2DCoop}(a), only the monostatic measurements of Agent 3 are
considered, illustrating the adverse effect of room symmetries and resulting unresolvable path overlap. In particular,
areas close to the walls are affected as well as the diagonals of the room. Fig.~\ref{fig:peb2DCoop}(b) shows the
information provided by the agents at $\bp\jnum{1}$ and $\bp\jnum{2}$ in their role as anchors. Their contribution is
similar to the fixed-anchor case analyzed in  Fig.~\ref{fig:peb2DToA}(c), but due to uncertainties in their own
positions, this information is not fully accessible. A robust, infrastructure-free positioning system is obtained if
these two components can complement one another. Indeed Fig.~\ref{fig:peb2DCoop}(c) indicates excellent performance
across the entire area. The distinction between the parts of the position-related information is further highlighted by
the CRLB ellipses in Fig.~\ref{fig:peb2DCoop}(c), which also include the fixed-anchor (bistatic) case of
Fig.~\ref{fig:peb2DToA}(c). It shows the decreased information of the cooperative part in comparison to the bistatic
case with fixed anchors. The monostatic ellipses are mostly oriented towards the nearest wall, where the most
significant information comes from. In many cases, this information is nicely complemented by the cooperative
contribution.

Fig. \ref{fig:pebCoopParts} shows the CDFs of the PEB in \eqref{eq:PEB_Coop} for $\Tp=\vu{1}{ns}$ and VAs of order one
and two. It is interesting to note that Multipath-Coop does not benefit from taking into account second-order MPCs. This
is explained by the large influence of the monostatic measurements, for which second-order reflections cause many
regions with unresolvable path overlap (c.f. Fig.~\ref{fig:peb2DCoop}(a)). For cooperative measurements, increasing the
VA order is still beneficial.

\begin{figure}[t]
  \centering
  \includegraphics[width=1\columnwidth]{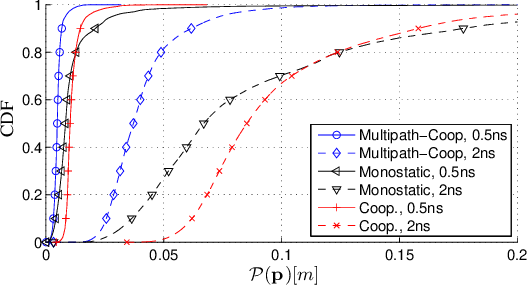}
  \caption{CDFs of the PEB in \eqref{eq:PEB_Coop} for Multipath-Coop with \vu{\Tp=0.5}{ns} and \vu{\Tp=2}{ns} for VAs of
  order two, showing contributions of different measurements types.} 
  \label{fig:pebcoopbw}
\end{figure}

Fig. \ref{fig:pebcoopbw} illustrates the influence of bandwidth on Multipath-Coop, using $\Tp=\vu{0.5}{ns}$ and
$\Tp=\vu{2}{ns}$ for VAs of order two. Especially for the monostatic measurements, the occurrence of unresolvable path
overlap is significantly reduced, leading to a clear advantage of a larger bandwidth.

\section{Conclusions and Outlook}
\label{sec:concl}
In this article, we have introduced and validated a unified framework for evaluating the accuracy  of radio-based
indoor-localization methods that exploit geometric information contained in deterministic multipath components. 
The analysis shows and quantifies fundamental
relationships between environment properties and the position-related information that can potentially be acquired. This
is due to two mechanisms: (i) Diffuse multipath, which is related to physical properties of  the propagation
environment, acts as interference to useful specular multipath components. (ii) Path overlap, which relates to system
design choices as the placement of agents but also to the given geometry of an environment, may render deterministic
components useless. An increased signal bandwidth allows to counteract those effects since it improves the
time-resolution of the measurements: The power of DM thus decreases and path overlap becomes less likely.

The framework allows for the analysis of different measurement setups: For instance, (i) in absence of
synchronization, position information can be extracted from the time-difference between MPCs. The need for clock-offset
estimation reduces thereby the positioning accuracy in comparison to a synchornized setup. (ii) Cooperation between
agents increases the available position-related information, but the uncertainty of the unknown positions of agents
\textit{acting }as anchors partly levels this effect. (iii) With monostatic measurements, the VAs move synchronously
with the agents, which leads to a scaling of the information provided by MPCs. These MPC-geometry-dependent scaling
factors lie between zero and two w.r.t. a conventional bistatic measurement. 

The quantification of position-related information, as provided by the presented framework, can be used for designing
positioning and tracking algorithms (e.g. \cite{MeissnerICC2014, LeitingerICC2014, Meissner2014a}). The proper
parametrization of the underlying geometric-stochastic channel model optimizes such algorithms and provides valuable
insight for system design choices such as antenna placements and signal parameters. Algorithms that can learn and
extract these environmental parameters online from measurements may achieve such optimization without the need for
manual system optimization and are thus an important topic for further research on robust indoor localization.

\appendices  
\numberwithin{equation}{section}
\section{FIM for orthogonal MPCs}
\label{app:lhf}
For a sampled received signal, the covariance matrix  of AWGN and the DM is written as 
\begin{align}
  \bCn = \sigman^2\IDMat_{\timestepsymub} + \bCc = \sigman^2\IDMat_{\timestepsymub} +
  \bar{\bS}\Herm\bS_\nu\bar{\bS}
\end{align}
where $\bar{\bS} = [\bs_{0}, \cdots , \bs_{\timestepsymub-1}]\T \in \mathbb{R}^{\timestepsymub \times
\timestepsymub}$ is the \emph{full} signal matrix with $\bs_i = \Big[ s((-i)\mod{N}\Ts),\dots,
s((\timestepsymub-1-i)\mod{N}\Ts) \Big]\T$, defined as a circulant matrix. The covariance matrix of DM is
\begin{align}
[\bar{\bS}\Herm\bS_\nu\bar{\bS}]_{n,m}
   = \sum_{i=0}^{\timestepsymub-1} \Ts S_\nu(i\Ts) &s((n-i)\mod{N}\Ts) \nonumber \\
&\times s((m-i)\mod{N}\Ts).
\end{align}
Using the Woodbury matrix identity, the inverse of  $\bCn$ can be written as
\begin{align}\label{eq:WoodburyInverseCov}
  \bCn^{-1} &= \frac{1}{\sigman^2} \Big[ \IDMat_\timestepsymub -
  \bar{\bS}\Herm\big(\sigman^2 \bS_\nu^{-1}+\bar{\bS}\bar{\bS}\Herm\big)^{-1}
  \bar{\bS} \Big] .
\end{align}
In (\ref{eq:sampledLHF}), this inverse is multiplied from the right by $\bS\balpha$, which can be re-written as
\begin{align*}
   \bCn^{-1}\bS\balpha &= \sum_{\mpcidxsym=1}^\mpcidxub \alpha\mpcidx\bCn^{-1}\bs_{\tau\mpcidx}
\\ &= \frac{1}{\sigman^2} \sum_{\mpcidxsym=1}^\mpcidxub \alpha\mpcidx\Big[ \IDMat_\timestepsymub -
  \bar{\bS}\Herm\big(\sigman^2 \bS_\nu^{-1}+\bar{\bS}\bar{\bS}\Herm\big)^{-1}
  \bar{\bS} \Big]\bs_{\tau\mpcidx} 
\end{align*}
where the factor $\bar{\bS}  \bs_{\tau_{\mpcidxsym}}$ on the very right is an autocorrelation vector 
of the transmitted signal shifted to delay time $\tau_\mpcidxsym$. The desired properties of $s(t)$---a large bandwidth
and favorable autocorrelation properties---imply that this autocorrelation has most of its energy concentrated at delay
$\tau_\mpcidxsym$. It hence samples the nonstationary PDP at  time $\tau_\mpcidxsym$ and we can replace $\bS_\nu$ for
each summand by a stationary PDP $\bS_\nu^{(\tau\mpcidx)} = \Ts S_\nu(\tau\mpcidx)\IDMat_\timestepsymub$.  Using this
assumption, we define  
\begin{align*}
  \big[\bCn^{(\tau\mpcidx)}\big]^{-1} &= 
\big[\sigman^2 \IDMat_\timestepsymub  +
 \Ts S_\nu(\tau\mpcidx) \bar{\bS}\Herm\bar{\bS} \big]^{-1}
\end{align*}
which involves the inverse of a cyclic matrix that can be diagonalized by a DFT. We introduce a unitary
DFT matrix $\bW$, $\bW\Herm \bW =\bW \bW \Herm =\IDMat $, and use $\bar{\bS} = \bW \widetilde{\bS}\bW\Herm $, 
where $\widetilde{\bS} = \diag (\sqrt{N} \bW \bs_0 )$ is a diagonal matrix containing the DFT of $\bs_0\T$ (the first row of $\bar{\bS}$),  
to obtain
\begin{align}\label{eq:WoodburyInverseCov_Freq}
  \big[\bC_\mathrm{n}^{(\tau\mpcidx)}\big]^{-1} 
&= 
\big[\bW \big(\sigman^2 \IDMat_\timestepsymub  +
 \Ts S_\nu(\tau\mpcidx) \widetilde{\bS}\Herm\widetilde{\bS} \big) \bW\Herm \big]^{-1}  \nonumber\\ 
&=
\bW \big(\sigman^2 \IDMat_\timestepsymub  +
 \Ts S_\nu(\tau\mpcidx) \widetilde{\bS}\Herm\widetilde{\bS} \big)^{-1} \bW\Herm  .
\end{align}
With this, we can approximate the second summand \label{eq:SS_LHFsampled} of likelihood function
(\ref{eq:sampledLHF}) by 
\begin{align*}
  [\balpha\Herm&\bS\Herm \bCn^{-1} \bS\balpha]_{\mpcidxsym,\mpcidxsym'} 
  \approx\alpha\mpcidx^*\alpha_{\mpcidxsym'}\bs_{\tau\mpcidx}\Herm \big[
  \bCn^{(\tau\mpcidx)}\big]^{-1} \bs_{\tau_{\mpcidxsym'}} \\
  & = \sum_{i=0}^{\timestepsymub-1} \frac{ \alpha\mpcidx^*\alpha_{\mpcidxsym'} |\mathrm{S}_f[i]|^2}{\sigman^2 + \Ts
  |\mathrm{S}_f[i]|^2 S_\nu(\tau\mpcidx)} \exp\left\{\frac{-j2\pi i (\tau\mpcidx -\tau_{\mpcidxsym'})}
  {\timestepsymub}\right\}
\end{align*}
where $\mathrm{S}_f[i]$ are samples of the DFT of $\bs_0$ and the exponential accounts for the delays $\tau\mpcidx$ and
$\tau_{\mpcidxsym'}$. Approximating the sum by an integral yields
\begin{align*}
  [\balpha\Herm&\bS\Herm \bCn^{-1} \bS\balpha]_{\mpcidxsym,\mpcidxsym'} \approx \\
  &\int_{f} \frac{\alpha\mpcidx^*\alpha_{\mpcidxsym'}|\mathrm{S}(f)|^2}{N_0 +
 |\mathrm{S}(f)|^2 S_\nu(\tau\mpcidx)} \exp\left\{-j2\pi f (\tau\mpcidx- \tau_{\mpcidxsym'})\right\}\mathrm{d}f.
\end{align*}
With this expression, the diagonal elements of submatrix $\bLA$ of the FIM can be written as
\begin{align} \label{eq:FIM_LA}
  [\bLA]_{\mpcidxsym,\mpcidxsym} &= 
  \E{\br|\bpsi}{ - \frac{\partial^2 \ln f(\br|\bpsi)}
  {\partial \tau\mpcidx \partial \tau\mpcidx}} \\
  &\approx 8\pi^2|\alpha\mpcidx|^2 \int_f f^2 \frac{|\mathrm{S}(f)|^2}{N_0 + S_\nu(\tau\mpcidx)|
  \mathrm{S}(f)|^2} \mathrm{d}f \nonumber \\
  &= \frac{8\pi^2}{N_0} \SINR\mpcidx \int_f f^2|\mathrm{S}(f)|^2\frac{N_0 + \Tp
  S_\nu(\tau\mpcidx)}{N_0 + |\mathrm{S}(f)|^2 S_\nu(\tau\mpcidx)}\mathrm{d}f \nonumber \\
  &= 8\pi^2\beta^2 \SINR\mpcidx \gamma\mpcidx \nonumber
\end{align}
where $\beta^2 = \int_f f^2 |S(f)|^2 \mathrm{d}f$ is the mean square bandwidth of $s(t)$, $\SINR\mpcidx =
|\alpha\mpcidx|^2/(N_0 + \Tp S_\nu(\tau\mpcidx))$ is the signal-to-interference-plus-noise ratio (SINR) of the
$\mpcidxsym$-th MPC, and $\gamma\mpcidx = \beta\mpcidx^2/\beta^2$ is called bandwidth extension factor, expressing the
influence of the whitening. The latter relates the mean square bandwidth of the \emph{whitened} pulse $\beta\mpcidx^2 =
\int_f f^2|\mathrm{S}(f)|^2 \frac{N_0 + \Tp S_\nu(\tau\mpcidx)}{N_0 + |\mathrm{S}(f)|^2 S_\nu(\tau\mpcidx)}\mathrm{d}f$
to $\beta^2$. Its value is a function of the interference-to-noise ratio $\Tp S_\nu(\tau\mpcidx) /N_0$. Note that $s(t)$
is assumed to be normalized to unit energy. Hence we have $|S(f)|^2 = \Tp$ for $|f| \le 1/(2\Tp)$ if $s(t)$ has a block
spectrum.

\section{Jacobian of VA Position w.r.t. Anchor Position}
\label{app:geometry}
\newcommand{\pva}{\bp\knum{\text{VA}}}
\newcommand{\pag}{\bp}
\newcommand{\qnum}[1]{\knum{#1}}
\newcommand{\pint}{\tilde{\bp}}

We want to find a simple expression for $\pa \bp\mpcidx\arb / \pa \bp\arb$. We restrict our derivation on a single
VA of a specific node w.l.o.g., so we drop all $\arbsymb,\mpcidxsym$-indexing and use a simpler notation $\pa \pva /
\pa \pag$. As explained in Section \ref{sec:Introduction}, $\pva$ is obtained by mirroring $\pag$ on walls $Q$ times
where $Q$ is the VA order. We use index $q$ for this iteration and refer to the intermediate positions as $\pint\q$
where $\pint\qnum{0} = \pag$ and $\pint\qnum{Q} = \pva$. We need to express $\pva$ as a function of $\pag$ and room
geometry. We account for the latter by considering walls with line equations
\begin{align}
  y-y\q = \tan(\zeta\q) (x - x\q)
\end{align}
where $\zeta\q$ is the wall angle and $\bd\q = (x\q,y\q)\T$ is an offset vector. We
obtain the $q$-th position by mirroring position $q-1$ on the $q$-th wall, or more formally 
\begin{align}
  \pint\q = \textbf{Mir}(\pint\qnum{q-1}, \zeta\q, \bd\q) \ .
\end{align}
where $\textbf{Mir}$ is defined as the mirroring operator. Starting at $q=Q$ and  using recursive substitution down to
$q=0$, we get
\begin{align}\label{eq:multimirroring}
  \pva =\textbf{Mir}( \ldots \textbf{Mir}( \textbf{Mir}(\pag, \zeta\qnum{1}, \bd\qnum{1}), \zeta\qnum{2},
  \bd\qnum{2}) \ldots, \zeta\qnum{Q}, \bd\qnum{Q}) \ .
\end{align}
The mirroring operation is given by
\begin{align}\label{eq:mirrorformula}
  \textbf{Mir}(\pint\qnum{q-1}, \zeta\q, \bd\q) 
  &= \bM(\zeta\q) (\pint\qnum{q-1} - \bd\q) + \bd\q \\
  &= \bM(\zeta\q) \pint\qnum{q-1} + \big(\bI - \bM(\zeta\q)\big) \bd\q \nonumber
\end{align}
where we use a mirror matrix that acts w.r.t. a line through the origin at angle $\zeta\q$,
\begin{align}
  \bM(\zeta\q) &=
  \left[ \begin{array}{lr}
  \cos(2\zeta\q) &  \sin(2\zeta\q) \\
  \sin(2\zeta\q) & -\cos(2\zeta\q) 
  \end{array} \right] \nonumber \\
  &= \textbf{Rot}(2\zeta\q) \flipmtx = \textbf{Rot}(2\zeta\q) \bF
\end{align}
and can be decomposed into a rotation by $2\zeta\q$, $\textbf{Rot}(2\zeta\q)$ and a sign-flip $\bF$ in the second
dimension. $\bM(\zeta\q)$ has eigenvalues $\{-1,+1\}$ and bears analogies to rotation. For breaking down
\eqref{eq:multimirroring}, we prefer the latter form of \eqref{eq:mirrorformula} because of the separated
$\pint\qnum{q-1}$-summand. By carefully repeated application, we obtain a formula
\begin{align}
  & \pva 
  = \bM(\zeta\qnum{Q}) \cdot \pint\qnum{Q-1} + \big(\bI - \bM(\zeta\qnum{Q})\big) \bd\qnum{Q} \nonumber \\
  &= \bM(\zeta\qnum{Q}) \bM(\zeta\qnum{Q-1}) \cdot \pint\qnum{Q-2} \ + \nonumber \\
  & \hspace{.48cm} \bM(\zeta\qnum{Q}) \big(\bI - \bM(\zeta\qnum{Q-1})\big) \bd\qnum{Q-1} + 
  \big(\bI - \bM(\zeta\qnum{Q})\big) \bd\qnum{Q} \nonumber \\
  & = \ \ldots \ = \bigg( \prod_{q=0}^{Q-1} \bM(\zeta\qnum{Q-q}) \bigg) \pag \ + \nonumber \\
  & \hspace{0.8cm} \sum_{q=1}^{Q} \bigg( \prod_{\tilde{q}=1}^{Q-q} \bM(\zeta\qnum{Q+1-\tilde{q}}) \bigg) \left(\bI -
  \bM(\zeta\q) \right) \bd\q
\end{align}
where the derivative w.r.t. $\pag$ is just the leading product of mirror matrices. Transposition reverses
multiplication order
\begin{align}\label{eq:gradientprod}
  \bigg( \f{\pa \pva}{\pa \pag} \bigg)\T = \prod_{q=1}^{Q} \bM(\zeta\q) \ .
\end{align}
To resolve this product, we derive a pseudo-homomorphism property of the mirror matrix. We note that both $\bF$ and
$\bM(\zeta)$ are symmetric, orthogonal, and self-inverse. Thus, $\bM(\zeta) = \textbf{Rot}(2\zeta)\bF$ implies
$\bM(\zeta)\bF =\textbf{Rot}(2\zeta)$. We rearrange the product of two mirror matrices
\begin{align*}
\bM(\zeta_a)\bM(\zeta_b) &= \bM(\zeta_a)\bM(\zeta_b)\T 
	= \textbf{Rot}(2\zeta_a)\bF \bF\T\textbf{Rot}(2\zeta_b)\T \\
 &= \textbf{Rot}(2\zeta_a)\bI \ \textbf{Rot}(-2\zeta_b) = \textbf{Rot}(2(\zeta_a-\zeta_b))
\end{align*}
and obtain the property
\begin{align}
  {\bf M}(\zeta_a) {\bf M}(\zeta_b) = {\bf M}(\zeta_a-\zeta_b) \bF \ .
  \label{eq:pseudohomo}
\end{align}
Applying \eqref{eq:pseudohomo} to \eqref{eq:gradientprod} $(Q-1)$-times the Jacobian of a VA position w.r.t. its
respective anchor's position yields
\begin{align}\label{eq:vatxjaco}
  \bigg( \f{\pa \pva}{\pa \pag} \bigg)\T 
  = \bM(\eff) \bF^{Q-1}	= \textbf{Rot}(2\eff) \bF^Q 
\end{align}
where we refer to $\eff := \zeta\qnum{1} - \zeta\qnum{2} + \ldots + (-1)^{Q-1} \zeta\qnum{Q} = \sum_{q=1}^{Q}
(-1)^{q-1} \zeta\q$ as the effective wall angle, where index $q$ iterates the order of occurrence of walls during MPC
reflection or VA construction.

\section{Delay Gradient for the Monostatic Setup}
\label{app:mono}
We transform the initial gradient from Appendix~\ref{app:geometry} into a magnitude-times-unit-vector form by
component-wise application of basic trigonometric identities. This yields an insightful expression for the monostatic
case, cf. \eqref{eq:monograd}. We consider
\begin{align*}
  & \be(\phi) - \be((-1)^{Q} \phi + 2\eff)
  =
  \left[ \begin{matrix} 
  \cos(\phi)-\cos((-1)^{Q} \phi + 2\eff)\\ 
  \sin(\phi)-\sin((-1)^{Q} \phi + 2\eff)
  \end{matrix} \right] \\ 
  &= \left[ \begin{matrix} 
  2\sin\left(\f{\left((-1)^{Q}+1\right)\phi + 2\eff}{2}\right)
  \sin\left(\hphantom{-}\f{\left((-1)^{Q}-1\right)\phi + 2\eff}{2}\right) \\ 
  2\cos\left(\f{\left((-1)^{Q}+1\right)\phi + 2\eff}{2}\right)
  \sin\left(-\f{\left((-1)^{Q}-1\right)\phi + 2\eff}{2}\right)
  \end{matrix} \right] .
\end{align*}
By defining symbols for the arguments that contain $\phi$ depending on the even/odd parity of Q
\begin{align*}
  & O := \f{(-1)^{Q}-1}{2}\phi + \eff 
  = \left\{ \begin{array}{ll}
    \eff & \ \text{If} \ Q \ \text{is even} \\
    \eff - \phi & \ \text{If} \ Q \ \text{is odd}
  \end{array} \right.\\
  & E := \f{(-1)^{Q}+1}{2}\phi + \eff
  = \left\{ \begin{array}{ll}
    \eff + \phi & \ \text{If} \ Q \ \text{is even} \\
    \eff & \ \text{If} \ Q \ \text{is odd}
  \end{array} \right.
\end{align*}
we further get
\begin{align}
  & \be(\phi) - \be((-1)^{Q} \phi + 2\eff)
  = 2 \sin(O) \ \be\Big(E-\f{\pi}{2}\Big) 
  \nonumber \\ & = 
  \left\{ \begin{array}{ll}
    2 \sin(\eff) \be(\phi + \eff - \f{\pi}{2}) 
    & \ \text{If} \ Q \ \text{is even} \\
    2 \sin(\eff - \phi) \be(\eff - \f{\pi}{2}) 
    & \ \text{If} \ Q \ \text{is odd}
  \end{array} \right. \ .
\end{align}

\section{Derivation of the NSync CRLB}
\label{app:TDOA}
\textit{Synchronized anchors:}
In order to derive the $3 \times 3$ EFIM $\FIM_{\bp, \clockoff}$ we need to repartition the transformation matrix
$\JacMat$ by combining the submatrices $\bH\anchoridx$ and $\bL\anchoridx = \bl_{\text{syn}}\anchoridx$ to $\bG\anchoridx = [\bH\anchoridx,
\bl_{\text{syn}}\anchoridx]$. Applying the transformation leads to 
\begin{align}\label{eq:TdoaFimTheta}
  &\FIM_\bP = \JacMat\T \FIM_\bpsi \JacMat = \\
  &\left[\! 
  \begin{array}{cccc}
    \sum\limits_{\anchoridxsym \in \mathcal{N}_\anchoridxsym}\!\big({\bG\anchoridx}\big)\T \!\bLA\anchoridx
    \bG\anchoridx\!&\! \big({\bG\jnum{1}}\big)\T\!\bLB\jnum{1}  \!&\!\!\cdots\! \!&\! 
    \big({\bG\jnum{\anchoridxub}}\big)\T\!\bLB\jnum{\anchoridxub} \\ 
    \big({\bLB\jnum{1}}\big)\T \bG\jnum{1} & \bLC\jnum{1} \!&\!   \!&\! \\
    \vdots \!&\!  \!&\!  \!\ddots\! &  \\
    \big({\bLB\jnum{\anchoridxub}}\big)\T \bG\jnum{\anchoridxub} \!&\!  \!&\!  \!&\! \bLC\jnum{\anchoridxub} 	
  \end{array}\!\right]. \nonumber
\end{align}
The $3\times 3$ EFIM is then given as the sum over the EFIMs of the corresponding anchors
\begin{align} \label{eq:Tdoa3x3EfimPO}
  &\FIM_{\bp, \clockoff} = \\ &
 \sum_{\anchoridxsym \in \mathcal{N}_\anchoridxsym}
  \big({\bG\anchoridx}\big)\T \Big[ \bLA\anchoridx - 
 \bLB\anchoridx \big(\bLC\anchoridx\big)^{-1} \big({\bLB\anchoridx}\big)\T \Big] \bG\anchoridx . \nonumber
\end{align}
When neglecting path overlap, this reduces to 
\begin{equation}
  \FIM_{\bp, \clockoff} = 
  \sum_{\anchoridxsym \in \mathcal{N}_\anchoridxsym}
  \big({\bG\anchoridx}\big)\T \bLA\anchoridx \bG\anchoridx,
\end{equation}
which leads finally to \eqref{eq:Tdoa3x3Efim}.

\textit{Asynchronous anchors:} The result for $\FIM_\btheta$ \eqref{eq:TdoaFimTheta} is also valid when considering
asynchronous anchors, provided that we respect $\bL\anchoridx = \bL_{\text{asyn}}\anchoridx$ and $\bG\anchoridx =
[\bH\anchoridx, \bL_{\text{asyn}}\anchoridx]$. We apply the blockwise inversion lemma twice, first to derive the EFIM
$\FIM_{\bp, \clockoffvec}$ (note that now $\clockoffvec$ is a vector), and then again to proof the additivity of the
EFIMs $\FIM_\bp\anchoridx$.

The EFIM $\FIM_{\bp, \clockoffvec}$ is now a square matrix of order $2+\anchoridxub$. It can be expressed as in
\eqref{eq:Tdoa3x3EfimPO}, but taking account of the changed definition of $\bG\anchoridx$. We can write its structure as
\begin{equation} \label{eq:TdoaExtendedEfimStructure}
  \FIM_{\bp, \clockoffvec} = 
  \sum_{\anchoridxsym \in \mathcal{N}_\anchoridxsym} 
  \begin{bmatrix}
    {\FIM_A\anchoridx} & \FIM_B\anchoridx \\
    \left(\FIM_B\anchoridx\right)\T & \FIM_D\anchoridx	
  \end{bmatrix},
\end{equation}
with $\FIM_A\anchoridx \in\mathbb{R}^{2 \times 2}$, $\FIM_B\anchoridx \in\mathbb{R}^{2 \times \anchoridxub}$ and
$\FIM_D\anchoridx \in\mathbb{R}^{\anchoridxub \times \anchoridxub}$. Further evaluation yields, that only the
$\anchoridxsym$-th column of $\FIM_B\anchoridx$ is nonzero, and the sum over $\FIM_B\anchoridx$ can be written as
\begin{equation}
	\sum_{\anchoridxsym \in \mathcal{N}_\anchoridxsym} \FIM_B\anchoridx 
	= \left[ \bb\jnum{1}, \dots, \bb\jnum{\anchoridxub} \right], \ \bb\anchoridx \in \mathbb{R}^2,
\end{equation}
meaning that each column is determined by the contribution of a different anchors.
Similarly, $\FIM_D\anchoridx$ has only one nonzero entry $\left[ \FIM_D\anchoridx \right]_{\anchoridxsym,
\anchoridxsym}$, leading to
\begin{equation}
  \sum_{\anchoridxsym \in \mathcal{N}_\anchoridxsym} \FIM_D\anchoridx = \text{diag}\left( \left[ \FIM_D\jnum{1}
  \right]_{1, 1}, \dots, \left[ \FIM_D\jnum{\anchoridxub} \right]_{\anchoridxub, \anchoridxub} \right).
\end{equation}
Rewriting $\FIM_{\bp, \clockoffvec}$ \eqref{eq:TdoaExtendedEfimStructure} and again applying the blockwise inversion
lemma yields the additivity of the EFIMs $\FIM_\bp\anchoridx$:
\begin{equation}
  \FIM_\bp = \sum_{\anchoridxsym \in \mathcal{N}_\anchoridxsym} 
  \FIM_A\anchoridx - \frac{1}{\left[ \FIM_D\anchoridx \right]_{\anchoridxsym, \anchoridxsym}} \bb\anchoridx
  \left(\bb\anchoridx\right)\T = \sum_{\anchoridxsym \in \mathcal{N}_\anchoridxsym} \FIM_\bp\anchoridx.
\end{equation}
The involved terms are defined by
\begin{equation*}
  \FIM_A\anchoridx = \big(\bH\anchoridx \big)\T \Big(\bLA\anchoridx -
  \bLB\anchoridx \big(\bLC\anchoridx \big)^{-1} \big(\bLB\anchoridx \big)\T \Big) \bH\anchoridx , 
\end{equation*}
\begin{equation*}
  \left[ \FIM_D\anchoridx \right]_{\anchoridxsym, \anchoridxsym} = \sum_{u=1}^{\mpcidxub\anchoridx} \sum_{v=1}^{\mpcidxub\anchoridx} 
  \big[ \bLA\anchoridx - \bLB\anchoridx \big(\bLC\anchoridx \big)^{-1} \big(\bLB\anchoridx \big)\T  \big]_{u,v} ,
\end{equation*}
and
\begin{align*}
  \bb\anchoridx &= \\
  &\big(\bH\anchoridx \big)\T \Big(\bLA\anchoridx -
  \bLB\anchoridx \big(\bLC\anchoridx \big)^{-1} \big(\bLB\anchoridx \big)\T \Big) [1 \dots 1]_{1 \times
  \mpcidxub\jnum{\anchoridxub}}\T .
\end{align*}

\section{Derivation of the Multipath-Coop CRLB}
\label{app:coop}
The EFIM for the cooperative setup is defined as
\begin{align*}
  \FIM_\bP = \bH\T \text{diag}\Big(\bLambda\jmnum{1}{1}, \ldots, \bLambda\jmnum{1}{\agentidxub},
 \ldots, \bLambda\jmnum{\agentidxub+\anchoridxub}{\agentidxub}\Big) \bH ,
\end{align*}
being of size $2M \times 2M $. It can be written with subblock $\bH$ from (\ref{eq:Jacobian_coop}) in the canonical form
(\ref{eq:FIM_coop}). Matrix $ \bLambda\jmnum{j}{m}$ is defined in (\ref{eq:efim_coop}). The canonical form decomposes
the EFIM $\FIM_\bP$ into contributions from independent transmissions inbetween the agents
or between agents and fixed anchors. Matrix $\FIM_\bP$ consists of the following subblocks for $\arbsymbtwo,\arbsymbtwo'
\in \mathcal{N}_\agentidxsym = \{1, \ldots, \agentidxub \}$,
\begin{multline}\label{eq:efimsum}
  \left[\FIM_\bP\right]_{2\times2}^{\arbsymbtwo,\arbsymbtwo'}  \\ =\sum_{\anchoridxsym \in 
(\mathcal{N}_\agentidxsym \cup\mathcal{N}_\anchoridxsym)}
\sum_{\agentidxsym \in \mathcal{N}_\agentidxsym}\big(\bH^{(\anchoridxsym,\arbsymbtwo,  \agentidxsym)} \big)\T
  \bLambda\anchoragentidx \bH^{(\anchoridxsym,\arbsymbtwo',\agentidxsym)}
\end{multline}
where $\bH^{(\anchoridxsym,\arbsymbtwo,  \agentidxsym)}$ stacks the spatial delay gradients (\ref{eq:gradresult}) as
defined in Section \ref{sec:geometry}. Considering that only summand $(\anchoridxsym,\agentidxsym)$ of
\eqref{eq:efimsum} contributes to a block, for which either index $\anchoridxsym$ or index $\agentidxsym$ equals
$\arbsymbtwo$ or $\arbsymbtwo'$, we get the following subblocks:
\subsubsection{Off-diagonal blocks $\arbsymbtwo \ne \arbsymbtwo'$} 
\begin{align*}
  \left[\FIM_\bP\right]_{2\times2}^{(\arbsymbtwo,\arbsymbtwo')} &=
  \big(\bH^{(\anchoridxsym,\arbsymbtwo,\agentidxsym)}\big)\T \bLambda\anchoragentidx
  \bH^{(\anchoridxsym,\arbsymbtwo',\agentidxsym)}\Big|_{\anchoridxsym = \arbsymbtwo,\agentidxsym =
  \arbsymbtwo'} \\              
  &\quad+ \big(\bH^{(\anchoridxsym,\arbsymbtwo,\agentidxsym)}\big)\T \bLambda\anchoragentidx  
  \bH^{(\anchoridxsym,\arbsymbtwo',\agentidxsym)} \Big|_{\anchoridxsym = \arbsymbtwo', \agentidxsym =
  \arbsymbtwo} \\
  &= \big(\GR^{(\arbsymbtwo,\arbsymbtwo')}\big)\T \bLambda^{(\arbsymbtwo,\arbsymbtwo')}
  \GT^{(\arbsymbtwo',\arbsymbtwo)} \\
  &\quad+ \big(\GT^{(\arbsymbtwo',\arbsymbtwo)}\big)\T \bLambda^{(\arbsymbtwo',\arbsymbtwo)}
  \GR^{(\arbsymbtwo',\arbsymbtwo)},
\end{align*}
using the definitions for  $\GR^{(\arbsymbtwo,\arbsymbtwo')}$ and  $\GT^{(\arbsymbtwo,\arbsymbtwo')}$ from
Section~\ref{sec:geometry_bistatic}. With $\GR^{(\arbsymbtwo,\arbsymbtwo')} = \GT^{(\arbsymbtwo',\arbsymbtwo)}$
(Section~\ref{sec:geometry_bistatic}) and $\bLambda^{(\arbsymbtwo,\arbsymbtwo')} =
\bLambda^{(\arbsymbtwo',\arbsymbtwo)}$ we get
\begin{equation}\label{eq:EFIM_offdiag}
  \left[\FIM_\bP\right]_{2\times2}^{(\arbsymbtwo,\arbsymbtwo')} = 2\FIMC \jmnum{\arbsymbtwo}{\arbsymbtwo'} =
  2\big(\GT^{(\arbsymbtwo',\arbsymbtwo)}\big)\T\bLambda^{(\arbsymbtwo',\arbsymbtwo)}
  \GT^{(\arbsymbtwo,\arbsymbtwo')}.
\end{equation}
\subsubsection{Diagonal blocks $\arbsymbtwo=\arbsymbtwo'$}
\begin{align*}
  \left[\FIM_\bP\right]_{2\times2}^{\arbsymbtwo,\arbsymbtwo} &=
  \big(\bH^{(\arbsymbtwo,\arbsymbtwo,\arbsymbtwo)}\big)\T
  \bLambda^{(\arbsymbtwo,\arbsymbtwo)}\bH^{(\arbsymbtwo,\arbsymbtwo,\arbsymbtwo)} \\ 
	      & \quad + \sum_{\substack{\anchoridxsym \in \mathcal{N}_\agentidxsym \backslash \{\arbsymbtwo\} \\
\agentidxsym =  
  \arbsymbtwo}} \big(\bH^{(\anchoridxsym,\arbsymbtwo,\agentidxsym)}\big)\T \bLambda\anchoragentidx   
  \bH^{(\anchoridxsym,\arbsymbtwo,\agentidxsym)} \\
  & \quad +  \sum_{\substack{\agentidxsym \in \mathcal{N}_\agentidxsym \backslash \{\arbsymbtwo\} \\
  \anchoridxsym =  \arbsymbtwo}} \big(\bH^{(\anchoridxsym,\arbsymbtwo,\agentidxsym)}\big)\T 
  \bLambda\anchoragentidx\bH^{(\anchoridxsym,\arbsymbtwo,\agentidxsym)} \\
  &\quad +  \sum_{\anchoridxsym \in \mathcal{N}_\anchoridxsym}
  \left(\bH^{(\anchoridxsym,\arbsymbtwo,\arbsymbtwo)}\right)\T \bLambda^{(\anchoridxsym,\arbsymbtwo)} 
  \bH^{(\anchoridxsym,\arbsymbtwo,\arbsymbtwo)}\\
  &= \big(\GM^{(\arbsymbtwo)}\big)\T\bLambda^{(\arbsymbtwo,\arbsymbtwo)} \GM^{(\arbsymbtwo)} \\
  & \quad + \sum_{\anchoridxsym \in \mathcal{N}_\agentidxsym \backslash \{\arbsymbtwo\}} 
  \big(\GT^{(\anchoridxsym,\arbsymbtwo)}\big)\T \bLambda^{(\anchoridxsym,\arbsymbtwo)}   
  \GT^{(\anchoridxsym,\arbsymbtwo)}\\
  & \quad + \sum_{\agentidxsym \in \mathcal{N}_\agentidxsym \backslash \{\arbsymbtwo\}} 
  \big(\GR^{(\arbsymbtwo,\agentidxsym)}\big)\T
  \bLambda^{(\arbsymbtwo,\agentidxsym)} \GR^{(\arbsymbtwo,\agentidxsym)} \\
  &\quad +  \sum_{\anchoridxsym \in \mathcal{N}_\anchoridxsym}
  \big(\GT^{(\anchoridxsym,\arbsymbtwo)}\big)\T \bLambda^{(\anchoridxsym,\arbsymbtwo)}  
  \GT^{(\anchoridxsym,\arbsymbtwo)}
\end{align*}
using again  $\GR^{(\arbsymbtwo,\arbsymbtwo')}$ and  $\GT^{(\arbsymbtwo,\arbsymbtwo')}$ from
Section~\ref{sec:geometry_bistatic} and $\GM^{(\arbsymbtwo)}$ from Section~\ref{sec:geometry_mono}. 
With $\GR^{(\arbsymbtwo,\agentidxsym)} = \GT^{(\agentidxsym,\arbsymbtwo)}$ and $\bLambda\anchoragentidx =
\bLambda^{(\agentidxsym,\anchoridxsym)}$ due to reciprocity, we get
\begin{align}\label{eq:EFIM_ondiag}
  \left[\FIM_\bP\right]_{2\times2}^{(\arbsymbtwo,\arbsymbtwo)} &= \FIMMo \jnum{\arbsymbtwo} \nonumber +
  2\sum_{\agentidxsym \in \mathcal{N}_\agentidxsym \backslash \{\arbsymbtwo\}} \FIMAg \jmnum{\agentidxsym}{\arbsymbtwo} 
+    
  \sum_{\anchoridxsym \in \mathcal{N}_\anchoridxsym}\FIMAn \jmnum{\anchoridxsym}{\arbsymbtwo} \\
  &= \FIMMo \jnum{\arbsymbtwo} + 2\FIMAg \jnum{\arbsymbtwo} + \FIMAn \jnum{\arbsymbtwo} 
\end{align}
which implicitly defines the contributions from monostatic measurements, bistatic measurements inbetween agents,
and bistatic measurements between agents and fixed anchors.

\balance

\IEEEtriggeratref{0}
\bibliographystyle{IEEEtran}
\bibliography{IEEEabrv,Bib_Jabref}

\begin{thebibliography}{10}
\providecommand{\url}[1]{#1}
\csname url@samestyle\endcsname
\providecommand{\newblock}{\relax}
\providecommand{\bibinfo}[2]{#2}
\providecommand{\BIBentrySTDinterwordspacing}{\spaceskip=0pt\relax}
\providecommand{\BIBentryALTinterwordstretchfactor}{4}
\providecommand{\BIBentryALTinterwordspacing}{\spaceskip=\fontdimen2\font plus
\BIBentryALTinterwordstretchfactor\fontdimen3\font minus
  \fontdimen4\font\relax}
\providecommand{\BIBforeignlanguage}[2]{{%
\expandafter\ifx\csname l@#1\endcsname\relax
\typeout{** WARNING: IEEEtran.bst: No hyphenation pattern has been}%
\typeout{** loaded for the language `#1'. Using the pattern for}%
\typeout{** the default language instead.}%
\else
\language=\csname l@#1\endcsname
\fi
#2}}
\providecommand{\BIBdecl}{\relax}
\BIBdecl

\bibitem{ShenJSAC2012}
Y.~Shen, S.~Mazuelas, and M.~Win, ``Network {Navigation: Theory and
  Interpretation},'' \emph{IEEE Journal on Selected Areas in Communications},
  2012.

\bibitem{Conti2014}
A.~Conti, D.~Dardari, M.~Guerra, L.~Mucchi, and M.~Win, ``Experimental
  {Characterization of Diversity Navigation},'' \emph{IEEE Systems Journal},
  2014.

\bibitem{MazuelasSTSP2009}
S.~Mazuelas, A.~Bahillo, R.~Lorenzo, P.~Fernandez, F.~Lago, E.~Garcia, J.~Blas,
  and E.~Abril, ``Robust {Indoor Positioning Provided by Real-Time RSSI Values
  in Unmodified WLAN Networks},'' \emph{IEEE Journal of Selected Topics in
  Signal Processing}, vol.~3, no.~5, pp. 821--831, Oct 2009.

\bibitem{FiccoTMC2014}
M.~Ficco, C.~Esposito, and A.~Napolitano, ``Calibrating indoor positioning
  systems with low efforts,'' \emph{IEEE Transactions on Mobile Computing},
  vol.~13, no.~4, pp. 737--751, April 2014.

\bibitem{MaranoJSAC2010}
S.~Marano~and, W.~Gifford, H.~Wymeersch, and M.~Win, ``N{LOS identification and
  mitigation for localization based on UWB experimental data},'' \emph{IEEE
  Journal on Selected Areas in Communications}, 2010.

\bibitem{WymeerschTC2012}
H.~Wymeersch, S.~Marano, W.~Gifford, and M.~Win, ``A {Machine Learning Approach
  to Ranging Error Mitigation for UWB Localization},'' \emph{IEEE Transactions
  on Communications}, 2012.

\bibitem{LuICC2013}
H.~Lu, S.~Mazuelas, and M.~Win, ``Ranging{ likelihood for wideband wireless
  localization},'' in \emph{IEEE International Conference on Communications
  (ICC)}, 2013.

\bibitem{WymeerschProc2009}
H.~Wymeersch, J.~Lien, and M.~Z. Win, ``Cooperative {Localization in Wireless
  Networks},'' \emph{Proceedings of the IEEE}, 2009.

\bibitem{ShenGlobecomm2009}
Y.~Shen and M.~Win, ``O{n the Use of Multipath Geometry for Wideband
  Cooperative Localization},'' in \emph{IEEE Global Telecommunications
  Conference (GLOBECOM)}, 2009.

\bibitem{ParhizkarICASSP2014}
R.~Parhizkar, I.~Dokmanic, and M.~Vetterli, ``Single-{C}hannel {I}ndoor
  {M}icrophone {L}ocalization,'' in \emph{39{t}h {I}nternational {C}onference
  on {A}coustics, {S}peech, and {S}ignal {P}rocessing}, 2014.

\bibitem{LeigsneringSPM2014}
M.~Leigsnering, M.~Amin, F.~Ahmad, and A.~Zoubir, ``Multipath {Exploitation and
  Suppression for SAR Imaging of Building Interiors: An overview of recent
  advances},'' \emph{IEEE Signal Processing Magazine}, 2014.

\bibitem{DokmanicPNAS2013}
I.~Dokmanic, R.~Parhizkar, A.~Walther, Y.~M. Lu, and M.~Vetterli, ``Acoustic
  {E}choes {R}eveal {R}oom {S}hape,'' \emph{Proceedings of the {N}ational
  {A}cademy of {S}ciences}, 2013.

\bibitem{MeissnerEUCAP2012}
P.~Meissner and K.~Witrisal, ``Analysis of {Position-Related Information in
  Measured UWB Indoor Channels},'' in \emph{6th European Conference on Antennas
  and Propagation (EuCAP)}, 2012.

\bibitem{ShenTIT2010}
Y.~Shen and M.~Win, ``Fundamental {Limits of Wideband Localization; Part {I}: A
  General Framework},'' \emph{IEEE Transactions on Information Theory}, 2010.

\bibitem{ShenTIT2010A}
Y.~Shen, H.~Wymeersch, and M.~Win, ``Fundamental {Limits of Wideband
  Localization - {Part II}: Cooperative Networks},'' \emph{IEEE Transactions on
  Information Theory}, 2010.

\bibitem{Qi2006}
Y.~Qi, H.~Kobayashi, and H.~Suda, ``Analysis of wireless geolocation in a
  non-line-of-sight environment,'' \emph{IEEE Transactions on Wireless
  Communications}, vol.~5, no.~3, pp. 672 -- 681, 2006.

\bibitem{GodrichTIT2010}
H.~Godrich, A.~Haimovich, and R.~Blum, ``Target {Localization Accuracy Gain in
  MIMO Radar-Based Systems},'' \emph{IEEE Transactions on Information Theory},
  vol.~56, no.~6, pp. 2783 --2803, {J}une 2010.

\bibitem{RichterVTC2005}
A.~Richter and R.~Thoma, ``Joint maximum likelihood estimation of specular
  paths and distributed diffuse scattering,'' in \emph{IEEE Vehicular
  Technology Conference, VTC 2005-Spring}, 2005.

\bibitem{MichelusiTSP2012}
N.~Michelusi, U.~Mitra, A.~Molisch, and M.~Zorzi, ``{UWB Sparse/Diffuse
  Channels, {Part I}: Channel Models and Bayesian Estimators},'' \emph{IEEE
  Transactions on Signal Processing}, 2012.

\bibitem{DecarliJSTSP2014}
N.~Decarli, F.~Guidi, and D.~Dardari, ``A {Novel Joint RFID and Radar Sensor
  Network for Passive Localization: Design and Performance Bounds},''
  \emph{IEEE Journal of Selected Topics in Signal Processing}, 2014.

\bibitem{SantosTWC2010a}
T.~Santos, F.~Tufvesson, and A.~Molisch, ``M{odeling the Ultra-Wideband Outdoor
  Channel: Model Specification and Validation},'' \emph{IEEE Transactions on
  Wireless Communications}, 2010.

\bibitem{KaredalTWC2007}
J.~Karedal, S.~Wyne, P.~Almers, F.~Tufvesson, and A.~Molisch, ``A{
  Measurement-Based Statistical Model for Industrial Ultra-Wideband
  Channels},'' \emph{IEEE Transactions on Wireless Communications}, 2007.

\bibitem{WitrisalICC2012}
K.~Witrisal and P.~Meissner, ``Performance bounds for multipath-assisted indoor
  navigation and tracking ({MINT}),'' in \emph{IEEE International Conference on
  Communications (ICC)}, 2012.

\bibitem{Meissner2014a}
P.~Meissner, E.~Leitinger, and K.~Witrisal, ``U{WB for Robust Indoor Tracking:
  Weighting of Multipath Components for Efficient Estimation},'' \emph{IEEE
  Wireless Communications Letters}, vol.~3, no.~5, pp. 501--504, Oct. 2014.

\bibitem{Borish1984}
J.~Borish, ``Extension of the image model to arbitrary polyhedra,'' \emph{The
  Journal of the Acoustical Society of America}, 1984.

\bibitem{Kunisch2003}
J.~Kunisch and J.~Pamp, ``An ultra-wideband space-variant multipath indoor
  radio channel model,'' in \emph{IEEE Conference on Ultra Wideband Systems and
  Technologies}, 2003.

\bibitem{CarboneJIM2013}
P.~Carbone, A.~Cazzorla, P.~Ferrari, A.~Flammini, A.~Moschitta, S.~Rinaldi,
  T.~Sauter, and E.~Sisinni, ``Low complexity uwb radios for precise wireless
  sensor network synchronization,'' \emph{Instrumentation and Measurement, IEEE
  Transactions on}, vol.~62, no.~9, pp. 2538--2548, Sept 2013.

\bibitem{FroehleICC2013}
M.~Froehle, E.~Leitinger, P.~Meissner, and K.~Witrisal, ``C{ooperative
  Multipath-Assisted Indoor Navigation and Tracking (Co-MINT) Using UWB
  Signals},'' in \emph{IEEE ICC 2013 Workshop on Advances in Network
  Localization and Navigation (ANLN)}, 2013.

\bibitem{MeissnerICC2014}
P.~Meissner, E.~Leitinger, M.~Lafer, and K.~Witrisal, ``Real{-Time
  Demonstration System for Multipath-Assisted Indoor Navigation and Tracking
  (MINT)},'' in \emph{IEEE ICC 2014 Workshop on Advances in Network
  Localization and Navigation (ANLN)}, 2014.

\bibitem{LeitingerICC2014}
E.~Leitinger, M.~Froehle, P.~Meissner, and K.~Witrisal, ``Multipath-{Assisted
  Maximum-Likelihood Indoor Positioning using UWB Signals},'' in \emph{IEEE ICC
  2014 Workshop on Advances in Network Localization and Navigation (ANLN)},
  2014.

\bibitem{MolischTPROC2009}
A.~Molisch, ``Ultra-wide-band propagation channels,'' \emph{Proceedings of the
  IEEE}, 2009.

\bibitem{Kay1993}
S.~Kay, \emph{Fundamentals of{ Statistical Signal Processing: Estimation
  Theory}}.\hskip 1em plus 0.5em minus 0.4em\relax Prentice Hall Signal
  Processing Series, 1993.

\bibitem{VanTrees1968}
H.~L. Van~Trees, \emph{Detection, {Estimation and Modulation, Part {I}}}.\hskip
  1em plus 0.5em minus 0.4em\relax Wiley Press, 1968.

\bibitem{SteinbockTAP2013}
G.~Steinb\"ock, T.~Pedersen, B.~Fleury, W.~Wang, and R.~Raulefs, ``Distance
  {Dependent Model for the Delay Power Spectrum of In-room Radio Channels},''
  \emph{IEEE Transactions on Antennas and Propagation}, vol.~61, no.~8, pp.
  4327--4340, Aug 2013.

\bibitem{MeissnerPhD2014}
P.~Meissner, ``Multipath-{Assisted Indoor Positioning},'' Ph.D. dissertation,
  Graz University of Technology, 2014.

\bibitem{MeasureMINT2013}
\BIBentryALTinterwordspacing
P.~Meissner, E.~Leitinger, M.~Lafer, and K.~Witrisal, ``M{easureMINT UWB
  database},'' 2014, {Publicly available database of UWB indoor channel
  measurements}. [Online]. Available:
  \url{www.spsc.tugraz.at/tools/UWBmeasurements}
\BIBentrySTDinterwordspacing

\end{thebibliography}


\begin{IEEEbiography}[{\includegraphics[width=1in,height=1.25in,clip,keepaspectratio]{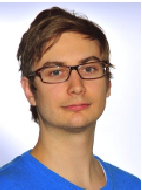}}]{Erik Leitinger} (S'12) was born in Graz, Austria, on March 27, 1985. He received the B.Sc. degree (with distinction) in electrical engineering from Graz University of Technology, Graz, Austria, in 2009, and the Dipl.-Ing. degree (with distinction) in electrical engineering from Graz University of Technology, Graz, Austria, in 2012.

He is currently pursuing his PhD degree at the Signal Processing and Speech Communication Laboratory (SPSC) of Graz University of Technology, Graz, Austria focused on UWB wireless communication, indoor-positioning, estimation theory, Bayesian inference and statistical signal processing.
\end{IEEEbiography}

\begin{IEEEbiography}[{\includegraphics[width=1in,height=1.25in,clip,keepaspectratio]{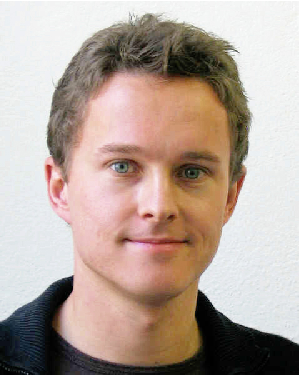}}]{Paul Meissner} (S'10--M'15) was born in Graz, Austria, in 1982. He received the B.Sc. and Dipl.-Ing. degree (with distinction) in information and computer engineering from Graz University of Technology, Graz, Austria in 2006 and 2009, respectively.  He received the Ph.D. degree in electrical engineering (with distinction) from the same university in 2014.

Paul is currently a postdoctoral researcher at the Signal Processing and Speech Communication Laboratory (SPSC) of Graz University of Technology, Graz, Austria. His research topics include statistical signal processing, localization, estimation theory and propagation channel modeling. He served in the TPC of the IEEE Workshop on Advances in Network Localization and Navigation (ANLN) at the IEEE Intern. Conf. on Communications (ICC) 2015 and of IEEE RFID 2015.
\end{IEEEbiography} 

\begin{IEEEbiography}[{\includegraphics[width=1in,height=1.25in,clip,keepaspectratio]{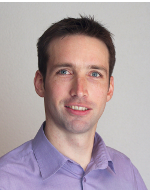}}]{Christoph
R\"udisser}  was born in Hohenems, Austria, in 1984. He received the
Dipl.-Ing. degree in electrical engineering (with distinction) from the
Graz University of Technology, Graz, Austria, in 2014.

Prior to his studies, he was at High Q Laser Production GmbH in Hohenems, Austria,
doing electronics and software development for four years. At present he is
looking for interesting career opportunities in the field of wireless
communications, with focus on statistical signal processing.
\end{IEEEbiography}

\begin{IEEEbiography}[{\includegraphics[width=1in,height=1.25in,clip,keepaspectratio]{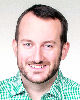}}]
{Gregor Dumphart} received the B.Sc. and Dipl.-Ing. degrees (with distinction) in information and computer engineering from Graz University of Technology, Graz, Austria in 2011 and 2014, respectively. 

He was a Student Assistant at the Department of Analysis and Computational Number Theory from 2009 to 2011 and at the Signal Processing and Speech Communication Laboratory from 2011 to 2013, both of Graz University of Technology.
Since October 2014, he is pursuing the PhD degree at the Communications Technology Laboratory, ETH Zurich, Zurich, Switzerland.
His research is concerned with localization and communication in dense networks (swarms) of low-complexity, sub-mm nodes by means of inductive coupling.	
\end{IEEEbiography}

\begin{IEEEbiography}[{\includegraphics[width=1in,height=1.25in,clip,keepaspectratio]{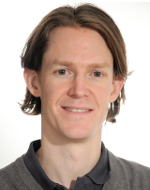}}]{Klaus Witrisal} (S'98--M'03) 
received the Dipl.-Ing. degree in electrical engineering from Graz University of Technology, Graz, Austria, in 1997 and the Ph.D. degree (cum laude) from Delft University of Technology, Delft, The Netherlands, in 2002.

He is currently an Associate Professor at the Signal Processing and Speech Communication Laboratory (SPSC) of Graz University of Technology, Graz, Austria, where he has been participating in various national and European research projects focused on UWB communications and positioning. 
He is co-chair of the Technical Working Group ``Indoor'' of the COST Action IC1004 ``Cooperative Smart Radio Communications for Green Smart Environments.'' His research interests are in signal processing for wideband and UWB wireless communications, propagation channel modeling, and positioning.

Prof. Witrisal served as a leading chair for the IEEE Workshop on Advances in Network Localization and Navigation (ANLN) at the IEEE Intern. Conf. on Communications (ICC) 2013, 2014, and 2015, as a TPC co-chair of the Workshop on Positioning, Navigation and Communication (WPNC) 2011, 2014, and 2015, and as a co-organizer of the Workshop on Localization in UHF RFID at the IEEE 5th Annual Intern. Conf. on RFID, 2011. He is an associate editor of IEEE Communications Letters since 2013. From 2007 to 2011, he was a co-chair of the MTT/COM Chapter of IEEE Austria.
\end{IEEEbiography}

\end{document}